\documentclass[aps,prx,twocolumn,superscriptaddress,noeprint,longbibliography, nofootinbib]{revtex4-2}

\usepackage{graphicx}
\usepackage{bm}
\usepackage{amsmath, amssymb}
\usepackage{booktabs}
\usepackage{mathrsfs}
\usepackage[normalem]{ulem}
\usepackage{pifont}
\newcommand{\cmark}{{\color{green!80!black} \ding{51}}}%
\newcommand{\xmark}{{\color{red!80!black}\ding{55}}}%
\usepackage[mathscr]{euscript}

\usepackage{physics}
\usepackage{xcolor}

\newcommand{\fsim}{\textsf{fSim}}
\newcommand{\modG}{\widetilde{G}}

\usepackage[colorlinks=true]{hyperref}
\hypersetup{citecolor = blue}

\begin{document}

\title{Many-body physics in the NISQ era: quantum programming a discrete time crystal}

\author{Matteo Ippoliti}
\affiliation{Department of Physics, Stanford University, Stanford, CA 94305, USA}

\author{Kostyantyn Kechedzhi}
\affiliation{Google Research, Venice, CA 90291, USA}

\author{Roderich Moessner}
\affiliation{Max-Planck-Institut f\"{u}r Physik komplexer Systeme, 01187 Dresden, Germany}

\author{S. L. Sondhi}
\affiliation{Department of Physics, Princeton University, Princeton, NJ 08540, USA}
\author{Vedika Khemani}
\affiliation{Department of Physics, Stanford University, Stanford, CA 94305, USA}

\date{\today}

\begin{abstract}
Recent progress in the realm of noisy, intermediate scale quantum (NISQ) devices~\cite{Preskill2018} represents an exciting opportunity for many-body physics by introducing new laboratory platforms with unprecedented control and measurement capabilities.
We explore the implications of NISQ platforms for many-body physics in a practical sense:
we ask which {\it physical  phenomena}, in the domain of quantum statistical mechanics, 
they may realize more readily than traditional experimental platforms. 
While a universal quantum computer can simulate any system, the eponymous noise inherent to NISQ devices practically favors certain simulation tasks over others in the near term. As a particularly well-suited target, we identify discrete time crystals (DTCs), novel non-equilibrium states of matter that break time translation symmetry. These can only be realized in the intrinsically out-of-equilibrium setting of periodically driven quantum systems stabilized by disorder induced many-body localization. While promising precursors of the DTC have been observed across a variety of experimental platforms - ranging from trapped ions to nitrogen vacancy centers to NMR crystals - 
none have \emph{all} the necessary ingredients for realizing a fully-fledged incarnation of this phase, and for detecting its signature long-range \emph{spatiotemporal order}. 
We show that a new generation of quantum simulators can be programmed to realize the DTC phase and to experimentally detect its dynamical properties, a task requiring extensive capabilities for programmability, initialization and read-out.
Specifically, the architecture of Google's Sycamore processor is a remarkably close match for the task at hand.
We also discuss the effects of environmental decoherence, and how they can be distinguished from `internal' decoherence coming from closed-system thermalization dynamics.
Already with existing technology and noise levels, we find that DTC spatiotemporal order would be observable over hundreds of periods, with parametric improvements to come as the hardware advances.

\end{abstract}

\maketitle


\section{Introduction}

The quest to build a universal quantum computer has fueled sustained progress towards the development of ``designer" many-body quantum systems 
across a variety of platforms ranging from trapped ions to superconducting qubits~\cite{GeorgescuRMP2014, QuantumSimulators2019}.
While the ultimate goal of a fault-tolerant quantum computer is still far into the future, the possibility of harnessing the computational power of the quantum world with noisy, intermediate scale quantum (NISQ)~\cite{Preskill2018} devices is already a reality. 
A notable milestone in this context was the recent announcement  of ``quantum supremacy'' (more accurately, ``quantum \emph{computational} supremacy''\footnote{Much of nature routinely carries out processes that are not simulable on a classical computer, but these are not recognizably computational tasks on highly controllable and thus recognizably computational devices. See 
Ref.~\onlinecite{AaronsonBlog} for a discussion of this point.}) in Google's Sycamore device, a solid-state, Josephson junction based platform with 53 qubits~\cite{Google2019}. 
While the computational task chosen for this purpose---simulating the output of random quantum circuits---may seem rather abstract and not useful in and of itself (though it does have at least one application~\cite{Aaronson}), a very active search for high-impact applications of NISQ devices is underway. In this vein, two recent works discussed how to implement highly structured circuits for quantum chemistry simulations~\cite{HartreeFock} and combinatorial optimization problems~\cite{QAOA} on Sycamore.

Now, a quantum computer is also necessarily a highly controllable many-body system~\cite{DiVincenzoCriteria}, and so these advances are also extremely tantalizing to many body physicists looking to push the frontiers of their own discipline. 
Indeed, Google's announcement, signifying a major breakthrough in computational science, also heralded the advent of a new \emph{laboratory system} with Hilbert spaces of significant size, which can potentially be used to host and discover new many-body physics.

This paper is motivated, broadly, by asking what the NISQ era of tunable, programmable quantum systems portends for many body physics;
and, narrowly, by asking what interesting physics could be realized immediately with Google's device.
Which \emph{physical phenomena} in the realm of quantum statistical mechanics can these devices realize, that have not yet been (as) crisply demonstrated in any other experimental setting?  As with the random circuit problem, a first demonstration should perhaps explore a landscape where some landmarks are already known and can be used to guide the search while leaving room for discovery.

Two conceptual challenges immediately present themselves to the many-body physicist: (i) The \emph{natural} time evolutions implemented on digital gate-based programmable simulators (such as Sycamore) are \emph{quantum circuits} rather than Hamiltonians. This is quite far from the typical setting in which condensed matter theory operates, which concerns the low-energy, long-wavelength emergent properties of equilibrium many-body systems. This is also distinct from regimes probed by analog simulators, such as cold-atom platforms, which generally target specific model Hamiltonians~\cite{BlochRMP2008, GrossBloch2017}.
And (ii) the tradeoffs between unitary control and platform size inevitably build some variation in individual circuit elements, which presents an additional challenge for simulating finely tuned model systems. 
We emphasize here that we are \emph{not} viewing these platforms as universal 
computational devices that can simulate any desired unitary evolution~\cite{Feynman1982, Lloyd1996, GeorgescuRMP2014} or allow computational investigation of the properties of particular Hamiltonians and quantum states~\cite{Wecker2015, McArdle2020, Rahmani2020}. 
Instead, due to near-term limitations in size and coherence time, we are interested in identifying physical phenomena that these platforms can \emph{immediately} and \emph{naturally} realize, as opposed to physics they could realize universally and asymptotically.

A parallel set of developments in quantum statistical mechanics furnishes a domain where these specific challenges turn into strengths: the study of non-equilibrium dynamics, and specifically the assignation of robust phase structure to many-body systems out of equilibrium. 
Remarkably, even without the conceptual framework of equilibrium thermodynamics, a possibility to identify phases and phase transitions remains~\cite{Sondhi2020,lpqo, Khemani2016}. 
This line of research has led to the discovery of new kinds of \emph{dynamical} many-body phenomena that may otherwise be forbidden by the strictures of equilibrium thermodynamics, with the \emph{discrete time crystal} (DTC) phase being the first and most paradigmatic example of this phenomenon~\cite{Khemani2016, Else2016, vonKeyserlingk2016, Khemani2019, SachaTCReview}.

Combining these insights leads us to focus on dynamical phases in disordered, out-of-equilibrium quantum matter -- specifically, many body localized (MBL) periodically driven (or Floquet) phases -- as natural 
candidates for the NISQ-era scientific program outlined above.  
Indeed, the quantum circuit structure which is Sycamore's {\it modus operandi} lends itself naturally to implement various Floquet drive protocols. 
Further, for these applications, randomness in circuit elements is not only tolerated, but is in fact necessary to stabilize the system against heating, and thus for observing interesting phenomena. 
For these reasons, in this work we propose precisely such a `physics-forward' use of the Sycamore device and its relatives: to realize an MBL Floquet DTC, a non-equilibrium many-body phase of matter that displays an entirely new form of \emph{spatiotemporal} order~\cite{Khemani2016, Else2016, vonKeyserlingk2016}. 
One striking feature of the DTC phase is that it spontaneously breaks the discrete time translation symmetry of the drive and exhibits \emph{period doubling}, a dynamical phenomenon with a long and rich history~\cite{faradaywaves, goldstein} which is has recently seen a resurgence in interest, with proposals spanning a wide range of classical and quantum systems~\cite{Holthaus1994, Sacha2015, Russomanno2017, Pal2018, Heugel2019, Kozin2019, YaoNayak2020}.

Our choice has several desirable aspects: 
(i) the DTC is a genuine collective \emph{many-body} phenomenon, and represents the best known example of a new paradigm in quantum statistical mechanics, that of an out-of-equilibrium phase of matter; 
(ii) it is of clear fundamental and conceptual importance, given its distinctive pattern of spatiotemporal order; and 
(iii) despite promising precursors~\cite{LukinExp, MonroeExp, BarrettExp, BarrettExp2}, a bona fide realization of this phase (or any many-body out-of-equilibrium phase, for that matter) has proved elusive for differing reasons in each of the existing experimental platforms in which it has been explored. 
Indeed, as we explain below, there are fundamental definitional aspects of the physics of this phase, specifically its central attributes of spatiotemporal order and robustness to choice of initial state, that have not yet been observed~\cite{Khemani2019}. 
Not only have these not been observed, detailed theoretical analysis has shown that these defining features are fundamentally \emph{absent} in the state-of-the-art experiments probing the DTC~\cite{Khemani2019}. Thus this proposal is \emph{not} about repeating previous experiments with incremental extensions to the scope of their observations; rather, it is about realizing and demonstrating the first genuine instance of this phase.

There is much reason to be optimistic.  The prior impressive experimental studies on DTCs have enabled a detailed understanding of the remaining obstacles to the realization of this phase, so that this goal appears eminently achievable in the near term. The resulting checklist contains several requirements that are hard to simultaneously satisfy in the previous setups. But these are sufficiently well-defined to be individually addressed and simultaneously realized on the Sycamore device. Indeed, as we show in this work, the \emph{existing} capabilities, architecture and gate-set in Sycamore satisfy all the desiderata, and the platform seems almost tailor made for this application!

We flesh out our proposal as follows. Sect.~\ref{sec:dtc_theory} contains a telegraphic account of the basics of DTCs to orient the following discussion. Sect.~\ref{sec:firstgen} presents a detailed account of the insights from previous experiments, from which we distill a list of experimental desiderata in Sect.~\ref{sec:ingredients} . Sec.~\ref{sec:nextgen} details how to meet these, and explains how to address the implementation of the required experimental protocol on a present-day quantum device, Google's Sycamore processor. 
We then provide evidence that the phenomenon we are looking for is indeed present for a range of experimentally achievable parameters (Sec.~\ref{sec:mbldtc}), and present an analysis of noise and other experimental imperfections to argue that its observation is possible despite present limitations of the NISQ platform (Sec.~\ref{sec:noise}). 
We conclude by discussing our results and directions for future work in Sec.~\ref{sec:discuss}.


\section{The discrete time crystal: Theory and Experiments}

We begin by briefly recapitulating the physics of the DTC phase in Sec.~\ref{sec:dtc_theory}, which defines the model and notation. This provides a minimal set of facts about the DTC needed to render this article self-contained; it may therefore be read diagonally by those with prior exposure to the field.
Sec.~\ref{sec:firstgen} discusses the state of the art in experimental efforts to engineer the DTC, followed in Sec.~\ref{sec:ingredients} by the enumeration of an experimental checklist of ingredients for realizing and observing this phase. These have not been simultaneously achievable in any single platform thus far. 
We refer the reader interested in an in-depth account of these issues to a review on time cystals by some of the present authors~\cite{Khemani2019}.

\subsection{Theoretical definitions and models}
\label{sec:dtc_theory}
 
 \subsubsection{The DTC and its import}
The canonical model of a (discrete) time crystal~\cite{Khemani2016} is realized in a Floquet system with a time-periodic Hamiltonian, with \emph{discrete} time-translation symmetry (dTTS) $H(t)= H(t+T)$. 
A DTC spontaneously breaks the dTTS of the drive: observables in this phase show periodic dynamics with a period $mT$, with $ \mathbb{Z}\ni m > 1$, corresponding to a sharp \emph{subharmonic} response in the frequency domain (for example, $m=2$ for period-doubled dynamics). 

The search for a time-crystal has roots dating back to age-old quests for perpetual motion machines, and this is a phase of matter that is provably disallowed by the strictures of equilibrium thermodynamics~\cite{Khemani2019, Watanabe_2020}. Hence, the intrinsically non-equilibrium setting of a periodically driven system is constitutive to realizing a time-crystal. 

Period doubling (or multiplexing) is ubiquitous in classical and quantum dynamical systems, in settings ranging from Faraday waves to parametric oscillators~\cite{faradaywaves, kuehn, goldstein}.
However these examples arise in single- or few-body systems, or in systems that are effectively few-body (in a mean-field sense)~\cite{Khemani2019}. On the other hand, defining a time crystal as a non-trivial, \emph{many-body} phase of matter requires us to consider macroscopic, strongly-interacting quantum systems. 
This is, in fact, the only setting in which time translation symmetry breaking is unexpected from the viewpoint of equilibrium thermodynamics; one- or few-body systems, such as simple harmonic oscillators, routinely exhibit oscillations and revivals in their dynamics.  

A pervasive challenge with periodically driven many-body systems is their tendency to absorb energy from the drive and thermalize to infinite temperature, maximizing entropy in the absence of conservation laws~\cite{PonteChandran_2015, Lazarides_2014}.
One robust mechanism for escaping this ``heat death" is many-body localization (MBL), wherein the dynamics fails to establish local thermal equilibrium even at arbitrarily late times due to disorder~\cite{Basko2006, Imbrie2016, Nandkishore2015, AbaninRMP2019, Lazarides_2015, PonteHuveneers_2015}. 
In particular, the system is thus prevented from heating to a trivial state.

A most striking property of this many-body localized phase is that it can now support new forms of order, which can be defined despite the inapplicability of the usual and familiar framework of equilibrium thermodynamics. The assignation of robust phase structure without relying on ground-states or equilibrium Gibbs states (or even time-independent Hamiltonians) is a fundamentally new paradigm in many-body physics, and the framework goes under the name of \emph{eigenstate order}~\cite{lpqo}. Most simply, many-body eigenstates of the system's Hamiltonian (or Floquet unitary) may \emph{individually} display non-trivial order and correlations, even as averages over eigenstates, such as in a Gibbs state, show no order! For example, the eigenstates may come in pairs, related to each other in the same way as the symmetry-broken ground states of a standard Ising ferromagnet. Unlike the latter, however, the pairing of states will be present \emph{throughout} the (quasi-)energy spectrum, with measurable dynamical consequences starting from states at all energies (instead of merely low-temperature ones). 

Eigenstate order of this type then underpins   
various non-trivial non-equilibrium phases, of both the symmetry-breaking and topological varieties. For a brief pedagogical introduction to nonequilibrium phase structure in Floquet systems, see Ref.~[\onlinecite{Moessner2017}].

In sum, the importance of the DTC is based on two pillars. First, it exhibits the spontaneous time-translation symmetry breaking expressed in its name, thereby closing out a centuries old quest for time-crystals and capturing the imagination of the general public.
Second, since such time-translational symmetry-breaking and spatiotemporal order is absent from {\it all} equilibrium phases, it stands out as, arguably, the most distinctive and striking instance of a new paradigm in many-body physics: an eigenstate-order based non-equilibrium phase of many-body matter.

 \subsubsection{Model Realizations}

We now turn to specific model realizations of this phase. A standard model of a Floquet DTC is an Ising model periodically ``kicked'' by a rotation about the $\hat{x}$ axis~\cite{Khemani2016}. The dynamics probed at `stroboscopic' times, $t=nT, \; n\in \mathbb{Z}$ are captured by studying the properties of the `Floquet unitary', which is the time-evolution operator over one period,
\begin{align}
U_F &= e^{-i g \sum_i X_i} e^{-i T (H_z + H_{\rm int})} \;, 
\label{eq:FloquetU}
\end{align}
where $T \equiv 1 $ is the drive period, $X_i$ ($Z_i$) denote spin-$1/2$ Pauli $x$ ($z$) operators on site $i$, $H_z = \sum_{i,j} J_{ij} Z_i Z_j$ is a diagonal Hamiltonian with Ising symmetry $P=\prod_i X_i$, and  $H_{\rm int}$ represents additional \emph{generic} interactions that may be present (examples include longitudinal fields $H_{\rm int} = \sum_i h_i Z_i$ or XY interactions $H_{\rm int} = \sum_{ij} J^{\perp}_{ij} [X_i X_j + Y_i Y_j]$). Localizing the system to prevent heating will require disorder in the couplings $J_{ij}$.

The model in Eq.~\eqref{eq:FloquetU} can potentially realize a discrete time-crystal phase in the regime $g = \frac{1}{2}(\pi-\epsilon)$, with $\epsilon$ sufficiently small. This represents an imperfect `$\pi$-pulse' \emph{i.e.} a nearly $180^{\circ}$ rotation about the $x$ axis. 
To understand the properties of the phase, consider first the limit $\epsilon= H_{\rm int}=0$. 
In this case, it easy to see that starting with a product state in the $\hat{z}$ basis, one action of the unitary enacts a perfect $180^\circ$ rotation and flips all spins; these are then flipped back under a second action of $U(T)$, thereby showing period doubled dynamics, $\langle Z_i(mT) \rangle = (-1)^m \langle Z_i(0)\rangle$. 

While the $\epsilon=H_{\rm int}=0$ limit is illustrative, defining the DTC as a \emph{phase} of matter requires some degree of stability to the choice of parameters and interactions. Indeed, what is remarkable is that under suitable conditions (requiring the presence of MBL), the dynamics can remain robustly \emph{locked at period doubling} for \emph{infinitely long} times in an \emph{extended region} of parameter space, \emph{i.e.} even for imperfect rotations ($\epsilon \neq 0$) and in the presence of generic perturbing interactions ($H_{\rm int}\neq 0$)~\cite{Khemani2016, Else2016, vonKeyserlingk2016}. 
We emphasize that this stability is inexplicable using any kind of semi-classical intuition; without quantum ordering, one would expect a finite deviation in rotation angle ($\epsilon\neq 0$) to accumulate over consecutive cycles, destroying the period doubling over a finite time scale $\sim \epsilon^{-1}$. 

Instead, the rigid locking of the dynamics to period doubling follows from the presence of long-range order \emph{in space} that stems from spontaneously breaking $\mathbb{Z}_2$ Ising symmetry,  whence `spatiotemporal' order~\cite{vonKeyserlingk2016}. This requires the Ising interactions $H_z$ to be the dominant part of the evolution during the first part of the drive. 
At any stroboscopic time, spins are locked into a ``frozen" pattern in space so that $\langle Z_i Z_j\rangle$ is nonzero for arbitrarily large $|i-j|$ even in highly-excited states (but can have a random, ``glassy" sequence of signs as a function of $i$, $j$). This pattern then flips every period. Notably, the DTC phase is also stable to the addition of interactions that \emph{explicitly} break Ising symmetry, such as longitudinal fields $H_{\rm int} = \sum_i h_i Z_i$~\cite{Else2016, vonKeyserlingk2016}. In this case, the long-range spatial order follows from spontaneously breaking an \emph{emergent} Ising symmetry~\cite{vonKeyserlingk2016}. 
This is a manifestation of the fact that the DTC phase is, in fact, stable to \emph{all} weak perturbations of the Floquet unitary \eqref{eq:FloquetU}, including those not encapsulated by $H_{\rm int}$ or $\epsilon$ --- a feature termed \emph{absolute stability} by a subset of the present authors~\cite{vonKeyserlingk2016}. 

In sum, the DTC is a robust, many-body phase of matter with spatiotemporal order (long-range order in space + infinitely long-lived period doubling dynamics in time), realized in the intrinsically non-equilibrium setting of periodically driven, MBL quantum systems. Probing spatiotemporal order requires measuring site-resolved spatial correlations, e.g. $\langle Z_i Z_j\rangle$, and temporal autocorrelation functions, e.g. $\langle Z_i(n) Z_i\rangle$.

\subsection{First Generation DTC Experiments}
\label{sec:firstgen}

The DTC phase is particularly amenable to experimental detection due to its stability and its distinctive measurable dynamical signatures. 
Indeed, the theoretical prediction of this phase was rapidly followed by a pair of experiments, one on disordered trapped ions in 1D~\cite{MonroeExp} and the other on disordered nitrogen vacancy (NV) centers in 3D diamond~\cite{LukinExp}. An experiment using nuclear magnetic resonance (NMR) on a clean crystalline 3D solid followed soon after~\cite{BarrettExp, BarrettExp2}. We will refer to this set of experiments as ``First Generation'' (FirstGen) time-crystal experiments.  

Each of the FirstGen experiments simulates a model drive captured by Equation~\eqref{eq:FloquetU}. The experiments differ in various key details and, between them, realize a varied matrix of parameters such as spatial dimension, range and type of interactions, nature of disorder, state preparation capabilities, microscopic controllability etc. 
Each one represents an experimental {\it tour de force}, and manages to observe temporal
signatures of DTC behavior (i.e. a signal locked at period doubling) over a finite extent in parameter space for the (finite) coherence time of the experiment. Despite the numerous differences between the platforms, the observed signatures look remarkably similar. 
However, despite these encouraging results, none of these platforms have all the ingredients needed for a genuine, asymptotic incarnation of the MBL DTC phase~\cite{Khemani2019}.

A key challenge for all three experiments lies in stabilizing MBL. 
Despite this, all three platforms still observe long-lived precursors of DTC order. 
This is because, even in cases where MBL is disallowed, it may nevertheless be possible to engineer a separation of scales such that thermalization happens on a parametrically \emph{slow} scale -- referred to as a `prethermal' regime in certain cases~\cite{Mori2016, Abanin2017,  Else2017, RubioAbadal2020, Cappellaro_Prethermal}. 
Specifically: the diamond NV center experiment~\cite{LukinExp} is incompatible with MBL because of its long-ranged interactions, but instead realizes a `critical TC' which thermalizes in a power-law slow fashion~\cite{CriticalTCPRL}.
Likewise, the NMR setup~\cite{BarrettExp} has no disorder and hence no MBL, and the long-lived signal therein was later explained as a prethermal phenomenon associated with a weakly broken global conservation law~\cite{Luitz2020}.
Finally, the trapped ion setup~\cite{MonroeExp} is the smallest and most controllable, and has many of the necessary ingredients for realizing MBL; 
however, it was shown in Ref.~\cite{Khemani2019} that, unexpectedly, the nature of disorder in the trapped ion TC experiment in Ref.~\cite{MonroeExp} is also \emph{not} sufficient for localization, and the signal observed therein also turned out to be of a prethermal rather than asymptotic nature. (However, as we discuss below, future iterations of the trapped ion experiment could, in principle, mitigate some of the issues of the first experiment). 

Despite not realizing an asymptotic MBL DTC, all three FirstGen experiments (and others~\cite{Pal2018}, mentioned below) have greatly advanced our conceptual understanding of the DTC phase and led to new theoretical insights. These include the elucidation of a new mechanism for prethermalization~\cite{Luitz2020} following the NMR experiment, and an understanding of the distinct types of disorder needed to stabilize MBL phases with distinct types of quantum order~\cite{Khemani2019}. These insights have enabled us to formulate a detailed checklist of desired experimental capabilities for the next generation of DTC experiments. As an example, the eventual theoretical understanding of the FirstGen experiments as prethermal (or slowly thermalizing) phenomena -- albeit of conceptually distinct genres -- emphasizes that a key experimental challenge is to distinguish a genuine MBL DTC phase from a transient prethermal version. We emphasize that is an issue because of the \emph{finite times} accessible to experiments rather than finite size (the diamond and NMR experiments have millions of spins so small systems sizes are not an issue, and slow prethermal dynamics stemming from large separations of parameter values arises even in infinitely large systems). In principle, the main difference between localized and prethermal DTCs lies in the lifetime of their quantum order: infinite for the former, transient for the latter. However, the ubiquity of environmental decoherence makes this distinction void in practice -- \emph{measured} DTC signals will be transient no matter what. Nevertheless, as we discuss below, fine-grained measurements of spatially resolved observables on a variety of initial states \emph{can} discriminate between prethermal and asymptotic TCs, even within finite experimental lifetimes.

\subsection{Experimental checklist}
\label{sec:ingredients}

In all, the FirstGen DTC experiments, with their varied strengths and limitations, have been instrumental in distilling a checklist of experimental ingredients needed for the realization and detection of a bona fide DTC phase. 
These ingredients, and their presence or absence in the various experiments, are summarized in Table~\ref{tab:summary} and articulated in more detail below; these serve to achieve two intertwined goals:
\begin{itemize}
    \item {\bf Realizing} a genuine \emph{asymptotic} MBL DTC phase, i.e. engineering all the theoretical criteria for achieving MBL and DTC order, so that an `ideal' experiment (without external decoherence) would observe an infinitely long-lived signal. 
    
    This is a matter of principle - if \emph{internal} decoherence (due to many-body quantum thermalization) in an ideal, noise-free incarnation of the platform destroys the signal at late times, then the system does \emph{not} realize an asymptotic DTC phase (this is true of all FirstGen experiments). On the other hand, if the lifetime is predominantly limited by external decoherence, then this is an issue of engineering that will see sustained improvement with future hardware innovations.
    
    \item {\bf Detecting} the spatiotemporal order that is a defining feature of the phase. This also entails experimentally discriminating between asymptotic (infinitely long-lived) and prethermal (transient) variants of DTCs, even within the constraints of environmental decoherence and finite experimental lifetimes. 
\end{itemize}

\begin{table}
    \centering
    \begin{tabular}{@{}lcccc@{}} \toprule
    Requirements & \multicolumn{4}{c}{Experiments} \\
    \cmidrule{2-5}
         & NV  & Trapped  & NMR  & Sycamore  \\ 
         & centers & ions & crystal &   \\ \midrule
         {\bf Definitional} & & & & \\
         $\quad$~Long coherence time & \cmark & \cmark & \cmark & \cmark \\ 
         $\quad$~Many-body & {\cmark\kern-1em\cmark} & $\boldsymbol{\sim}$ & {\cmark\kern-1em\cmark} & \cmark \\ 
         {\bf Stabilizing MBL} & & & & \\
         $\quad$~Short-range int. & \xmark & {\bf ?} & \xmark & \cmark \\
         $\quad$~Ising-even disorder & \cmark & \xmark & \xmark & \cmark \\
         {\bf Detection} & & & & \\
         $\quad$~Site-resolved meas. & \xmark & \cmark & \xmark & \cmark \\
         $\quad$~Varied initial states & \xmark & $\boldsymbol{\sim}$ & \xmark & \cmark \\
         \bottomrule
    \end{tabular}
    \caption{Summary of experimental requirements for realizing and observing DTC spatiotemporal order, and the relative merits of different experimental platforms. The `double' check-marks for the NV and NMR platforms in the `many-body' category are to emphasize that these setups, with $>O(10^6)$ constituents, are operating in the thermodynamic regime, at a size that is orders of magnitude larger than the trapped ion experiment ($\sim 10$ ions) and Sycamore ($\sim 50$ qubits).
    } 
    \label{tab:summary}
\end{table}

We now enumerate six desired experimental capabilities, grouped in three broad categories. 

\subsubsection{Basic definitional requirements}
As mentioned earlier, a DTC phase is characterized by infinitely long lived, quantum-coherent oscillations in infinitely large, macroscopic many-body systems. While an actual experiment will always be of finite size with a finite coherence time, non-trivial realizations still require both of these to be sizeable, with room for parametric improvements with engineering advances. Thus two basic requirements on the platforms are:
\vspace{6pt}

\noindent {(i) \bf Truly many-body.} 
The experimental systems should contain a number of qubits that does not qualify as ``few-body''. While there is no sharp boundary between ``few'' and ``many'', it is clear that the NV and NMR experiments satisfy this requirement ($>10^6$ qubits), while the trapped ion experiment (10 qubits) may be considered border-line -- a few tens to hundreds of qubits would more comfortably fit the description. 
An added bonus is if the platform permits one to vary the system size, which would allow for finite-size scaling analysis of various order parameters. 
Another scenario ruled out by this requirement is that of \emph{effectively few-body} systems where, despite a nominally large number of qubits, the dynamics becomes few-body in a mean-field sense. Several recent TC experiments fall in this category~\cite{Pal2018, Heugel2019, Kessler2020, Zhang2017}, with Ref.~\cite{Pal2018} furnishing a particularly nice example using NMR on `star-shaped' molecules. 
We remark that this point is \emph{not} about classical simulability, but specifically about physics. Time-crystals are only non-trivial for macroscopic many-body systems; few-body systems exhibit special phenomena (e.g. recurrences) that do not scale to the many-body limit, and could prove confounding to the observation of the desired phenomenon. 
\vspace{6pt}

\noindent {(ii) \bf Long Coherence time.}
Experimental platforms aiming to exhibit dynamical phases clearly must be able to preserve quantum coherence for long enough, so that the underlying dynamical phenomena can be distinguished from short-time transients. Again, while there is no sharp boundary, revealing DTC order requires a coherence time of at least multiple tens to hundreds of Floquet cycles. We caution, however, that this may still not be enough to discriminate between MBL and prethermal TCs without using additional fine-grained probes (cf. points (v) and (vi) below). 
All the FirstGen platforms had a lifetime on the order of 100 Floquet periods.

\subsubsection{Requirements for stabilizing MBL}
MBL is an essential ingredient for realizing a robust DTC phase in an extended region of parameter space, and in preventing periodic driving from heating the interacting system to infinite temperature. However, MBL is only stable under certain conditions sensitive to the range of interactions, and the scope for engineering disorder:
\vspace{6pt}

\noindent {(iii) \bf Short-ranged interactions.} 
Long-ranged interactions are known to destabilize localization~\cite{FleishmanAnderson, Burin2015, Yaodipoles}.
Interactions with strength scaling as $1/r_{ij}^\alpha$ are incompatible with MBL if $\alpha<d$, where $d$ is the dimension of the system~\cite{FleishmanAnderson}.
They are perturbatively compatible with MBL\footnote{Note that for the purpose of this article, we are not concerning ourselves with the open question of possible \emph{non-perturbative} instabilities of MBL that may asymptotically destabilize localization in dimensions greater than one, or with power-law decaying interactions with any power~\cite{deroeckAvalanche}. These effects, if they exist, will happen for system sizes and time scales that are well beyond the capabilities of any near-term simulators.} if $\alpha>2d$~\cite{Burin2015}. 
Finally the regime $d<\alpha<2d$ is not fully understood in general; localization (or its absence) in this regime depends on the particular form of interactions present in specific Hamiltonians~\cite{Burin2015, Yaodipoles}. 
Out of the FirstGen experiments, the only one possibly satisfying the requirement of short-ranged interactions is the one based on trapped ions ($d=1$, $\alpha \approx 1.51$), but it is not presently settled whether the long-range Ising interactions therein are compatible with localization for the chosen value of $\alpha\approx 1.51$.\footnote{We thank S. Gopalakrishnan for a discussion of this point.}
This is indicated in Table~\ref{tab:summary} by a question mark. However we note that, in principle, $\alpha$ is a tunable parameter in the trapped ion platform, and hence the trapped ion experiment could be repeated in the future with a value of $\alpha > 2d$. 
Both the diamond NV and NMR experiments have $d = \alpha = 3$ and are thus not compatible with MBL. 
\vspace{6pt}

\noindent (iv) {\bf Dominantly Ising interactions with Ising-even disorder.}
While stabilizing MBL generically requires disorder in the drive parameters, the \emph{nature} of the disorder required to stabilize an MBL DTC is more specific: one requires strong disorder in \emph{Ising-even} interactions $H_z = \sum_{ij} J_{ij} Z_i Z_j$~\cite{Khemani2019} in a drive with dominantly Ising interactions of the form \eqref{eq:FloquetU}. If, instead, the only operators coupled to disorder are odd under the Ising symmetry $P_x = \prod_i X_i$ (as is the case e.g. for on-site fields $H_{\rm int} = h_i Z_i$), then this is not sufficient to stabilize MBL. This is because the Floquet evolution over two cycles, $U_F^2$, is only weakly disordered, and the dynamics is consequently not MBL. The effective disorder strength is weak because the Ising-odd disordered fields are `echoed out' by the approximate $\pi$-pulse, to leading order (see Appendix~\ref{app:echo_out} for a discussion of this point).
Of the FirstGen experiments, only the NV platform realizes Ising-even disorder due to the random position of NV centers in three-dimensional space; while this alone is not enough for MBL (because of the long-range interactions), the disorder still leads to algebraically slow themalization, giving a `critical time crystal' in the NV setup. 
The NMR system is clean and spatially ordered, and hence not localized.
Finally, the trapped ion experiment features disorder only in Ising-odd longitudinal fields, while the Ising-even interactions are non-random and well approximated as $J_{ij} \sim J_0/r_{ij}^\alpha$. 
In a finite lattice the displacements $r_{ij}$ of the trapped ions (and thus the interactions $J_{ij}$) will include weak inhomogeneities due to the interplay of Coulomb interactions with  
the confining trap; however these inhomogeneities are perfectly deterministic and reflection-symmetric, and turn out to be insufficient to stabilize MBL~\cite{Khemani2019}. 
In general, it is easier for many experimental setups to implement disorder in onsite fields rather than Ising couplings, and this requirement is a key engineering obstacle towards realizing DTCs on many such platforms, including in trapped ions.

\subsubsection{Requirements for detection}
Finally, we turn to the requirements of unambiguously demonstrating the asymptotic DTC phase and distinguishing it from its transient prethermal cousins -- \emph{even within the reality of finite experimental lifetimes.} 

\begin{figure}
    \centering
    \includegraphics[width=\columnwidth]{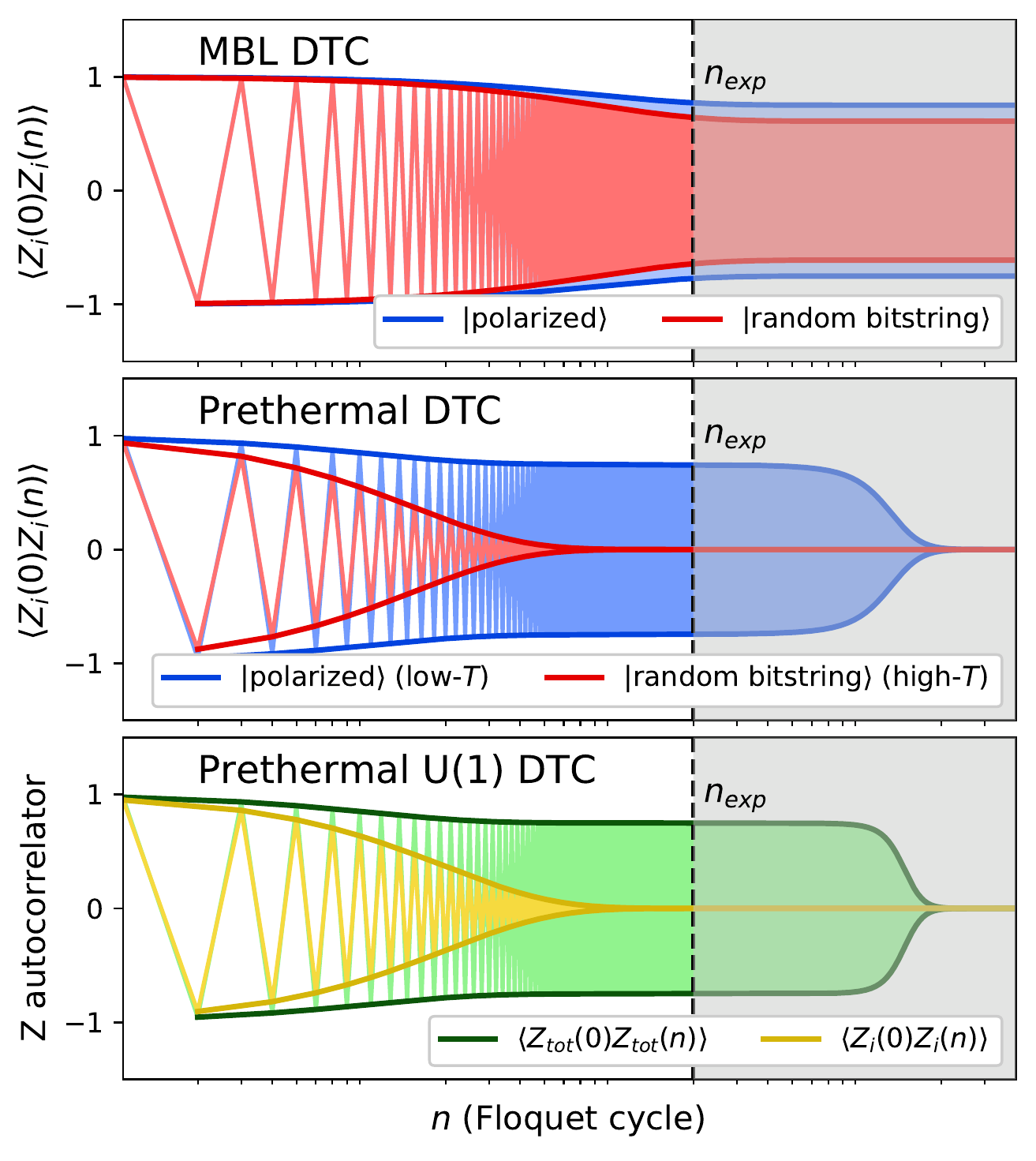}
    \caption{Illustrative sketch for distinguishing MBL and prethermal DTCs, even with access to only finite experimental coherence times (vertical dashed line, $n_\text{exp}$). 
    Top: in an MBL DTC, local autocorrelators remain large for all initial product states in the $Z$ basis.
    Center: in a prethermal DTC, low-temperature initial states have long-lived autocorrelators whose lifetime may exceed the experimental coherence time. However, generic bit-string initial states (high-temperature) decay quickly: the dependence on initial states -- visible within experimental time scales $n_\text{exp}$ -- is a signature of prethermalization. 
    Bottom: in a $U(1)$ prethermal DTC, the total magnetization $Z_\text{tot} = \sum_i Z_i$ is nearly conserved (for all initial states, including high-temperature ones). However local operators $Z_i$ decay quickly: the $U(1)$-prethermal behavior is revealed through site-resolved measurements. 
    }
    \label{fig:prethermal_sketch}
\end{figure}

The key discriminator is that MBL TCs show period doubled oscillations from \emph{all} generic short-range correlated initial states, while prethermal TCs only show long-lived oscillations from certain special initial states. In addition, spatiotemporally resolved correlators show long-range order and period doubling in MBL TCs, while certain variants of prethermal TCs only show long-lived oscillations in globally averaged observables but not site-resolved ones. Thus, studying varied initial states and making site-resolved measurements \emph{even for finite experimental times} would distinguish between an asymptotic MBL DTC and all known alternate mechanisms that could support a prethermal DTC, as illustrated in Fig.~\ref{fig:prethermal_sketch}. 

In more detail,the key idea of prethermal dynamics is that, in a suitable reference frame, the system behaves for a long time \emph{as though} it was governed by a static effective Hamiltonian (although the temperature of the state slowly increases en route to infinite temperature)~\cite{Mori2016, Abanin2017}.
If the effective Hamiltonian has an ordered phase below a critical temperature $T_c$, then a low-energy initial state would display quantum order for a long time, before eventually heating past $T_c$ thus causing the order to melt~\cite{Else2017}. 
However, a high-energy initial state would not show any order, even for short times. Thus, practically, a useful discriminatory criterion is the dependence of the signal on the choice of initial state. In MBL DTCs there should be no strong dependence (as the whole spectrum is localized). On the other hand, prethermal DTCs  associated with symmetry breaking display long-lived oscillations for low-temperature ordered states but not for others (Fig.~\ref{fig:prethermal_sketch}(a,b)). 

Separately, another mechanism for prethermalization is the emergence of a quasi-conserved quantity associated to an approximate symmetry of the prethermal Hamiltonian~\cite{Luitz2020, Abanin2017}. This mechanism for slow thermalization can be at play even for very high-temperature initial states. 
In this case, measurements of global observables such as the total magnetization are at risk of detecting the slow relaxation of a quasi-conserved quantity rather than the DTC pattern of spatiotemporal order. However, measuring site-resolved correlations can distinguish between prethermal U(1) DTCs and MBL DTCs (Fig.~\ref{fig:prethermal_sketch}(a,c))

Thus, one requires:
\vspace{6pt}

\noindent (v) {\bf Widely tunable initial states.} 
To distinguish localized and prethermal DTCs within a finite experimental lifetime, one needs to test a variety of initial states (prethermal DTCs are highly sensitive to the choice unlike MBL DTCs).
In practice, the ability to prepare any \emph{computational basis} state, i.e. product states in the $z$ basis, would be enough. 
This cannot be done on platforms that only allow for the preparation of special initial states, such as fully polarized ones. 
Of the FirstGen platforms, only the trapped ion experiment has the capability to widely vary initial states, although this was not fully explored in Ref.~\cite{MonroeExp}.  The experiment only considered two initial states: a fully polarized state, $|0\rangle^{\otimes L}$,  and a state polarized on the left and right halves, $|0\rangle^{\otimes L/2}|1\rangle^{\otimes L/2}$. However polarized or near polarized states are maximally ineffectual at distinguishing between MBL and prethermal dynamics~\cite{Khemani2019, Luitz2020}. Because the trapped ion experiment has long-range interactions, the effective Hamiltonian governing the prethermal dynamics can have an Ising symmetry breaking transition at a finite temperature $T_c$ even in one dimension, and near polarized states are in the low-temperature sector of the effective Hamiltonian. Indeed, detailed numerical simulations of the trapped ion experiment on a wider class of initial states found strong initial state dependence, with the DTC signal decaying much more rapidly for randomly picked computational-basis states, consistent with prethermal DTC order~\cite{Khemani2019}. Separately, a different mechanism for prethermalization entails the long-lived quasi-conservation of a global operator such as the total magnetization. Again, polarized initial states have large total magnetization and can show slow dynamics due to the quasi-conservation law, while randomly picked $z$ product states would not.  
\vspace{6pt}

\noindent (vi) {\bf Site-resolved measurements.}
Detecting genuine spatiotemporal order requires measuring site-resolved spatial correlation functions of the form $\langle Z_i Z_j \rangle$, in addition to temporal autocorrelators. 
This capability to locally probe individual qubits is also necessary for distinguishing MBL TCs from prethermal variants involving global quasi-conservation laws. For instance, the NMR experiment operates in an extremely \emph{hot} regime, with very high temperature initial states that would be well above the ordering temperature $T_c$ of the effective Hamiltonian; but these can still show slow dynamics in global observables that couple to a quasi-conservation law, such as the total magnetization~\cite{Luitz2020}. In contrast, local autocorrelators would show a fast decay in this regime. In contrast, site-resolved autocorrelators show oscillations forever in an MBL TC. 
The NMR and NV center experiments (which involve $>10^6$ qubits) are limited to probing spatially averaged quantities such as the total magnetization $\sum_i Z_i$, which do not provide the necessary resolution. Among the FirstGen experiments, only the trapped ion experiment satisfies this requirement. Table I summarizes the matrix of experimental desiderata and their availability in different FirstGen experiments.

We now turn to how the next generation (NextGen) of quantum simulators - such as the already operational Google Sycamore processor  - can be programmed to realize all these ingredients in turn, and hence to furnish the first bona fide realization of the time-crystal phase. We should note that while the trapped ion experiment has not yet demonstrated an MBL DTC phase, it may be possible for future iterations of this platform to do so. The key engineering challenges entail scaling up the system to suitably larger numbers of ions, and adding uncorrelated disorder in the Ising couplings $J_{ij}$ (which is possible, in principle, with extensively many tuning knobs~\cite{Lin2011, Korenblit2012}). 
These are achievable given enough time and effort. Likewise, quantum simulators using Rydberg or dressed Rydberg atoms meet almost all the desired criteria, and are currently limited only by their coherence time~\cite{Zeiher2016}.
Future improvements will no doubt also enable the observation of such phenomena on this versatile platform. 
However, as we demonstrate next, \emph{currently} existing capabilities in the Sycamore device already satisfy all the desiderata and, indeed, the platform seems tailor made for this application!


\section{Next Generation: Realizing a DTC on the Sycamore processor}
\label{sec:nextgen}

\begin{figure}
\centering
\includegraphics[width=0.26\columnwidth]{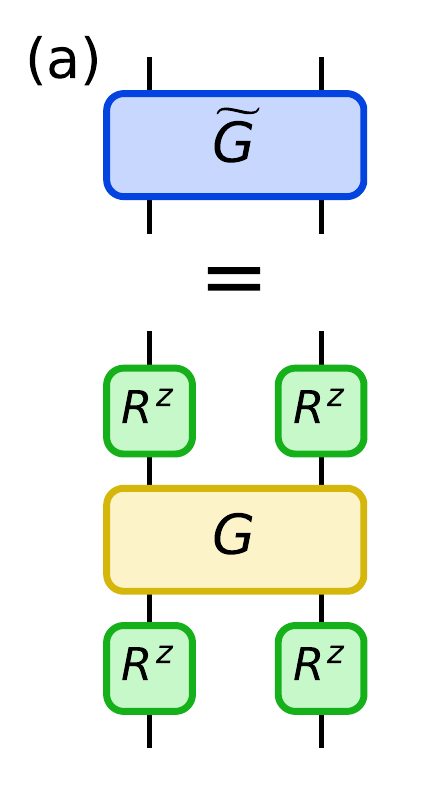}~
\includegraphics[width=0.7\columnwidth]{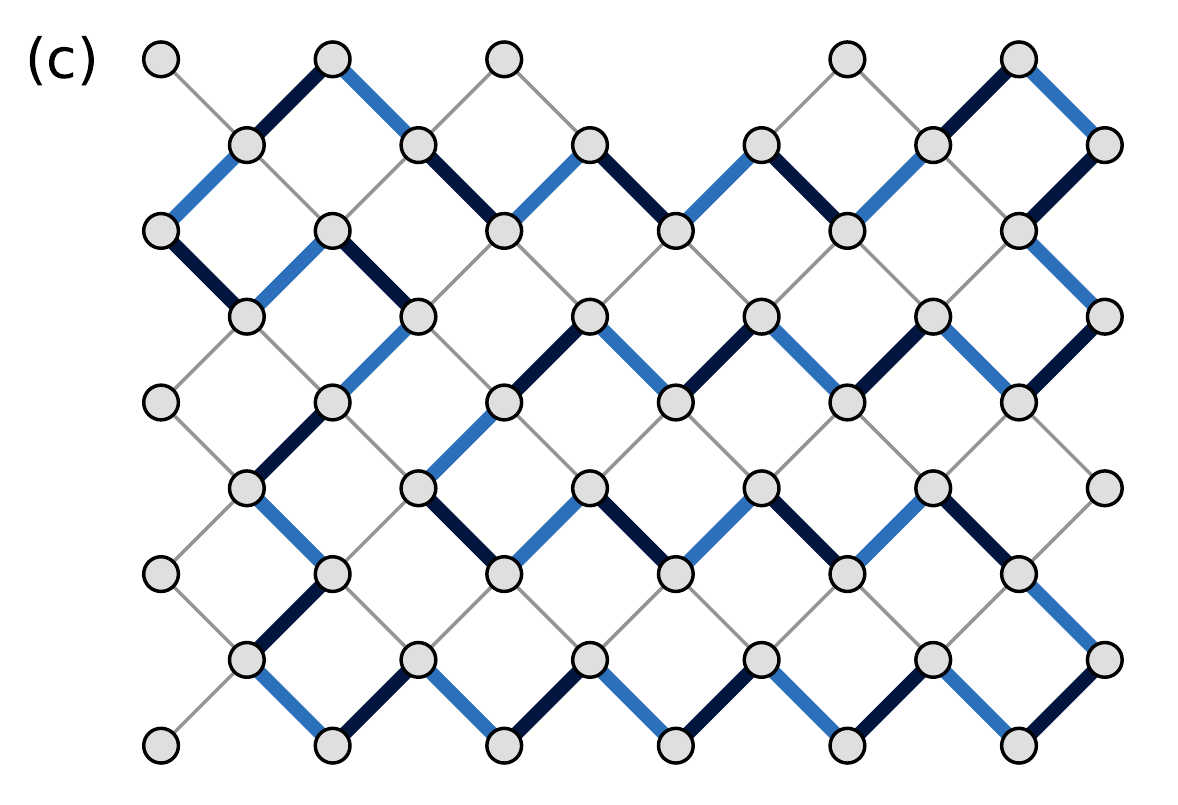}\\
\includegraphics[width=\columnwidth]{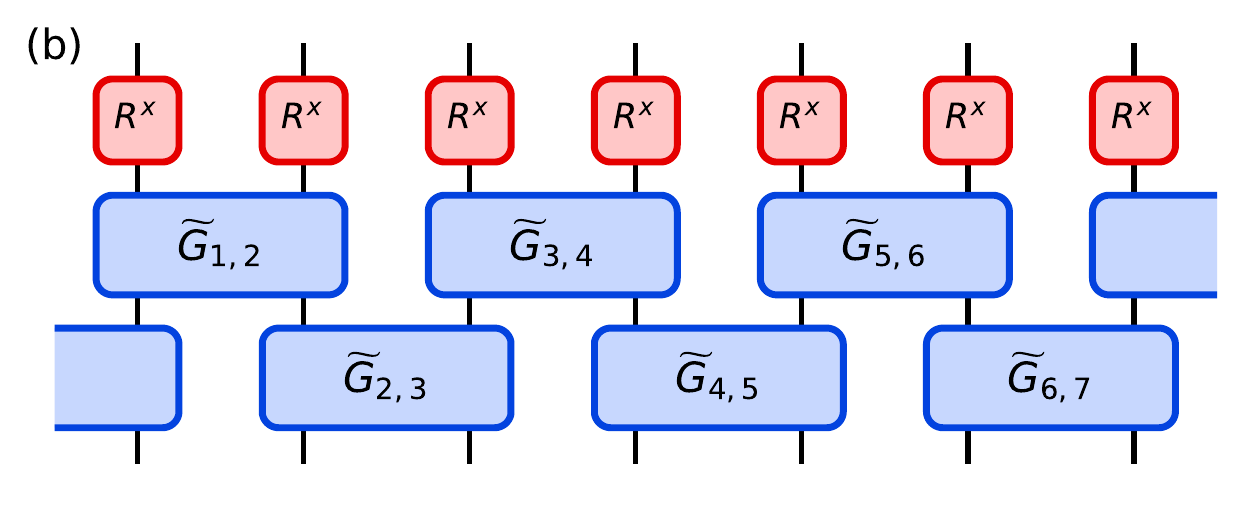}
\caption{Simulating a 1D Floquet DTC on the Sycamore chip.
(a) Modified gate $\modG$ in terms of the native gate $G$ and single-qubit $Z$ rotations.
(b) Circuit for the DTC Floquet unitary: each Floquet cycle acts with $\modG$ on each pair of neighboring qubits, followed by single-qubit $X$ rotations, as depicted.
(c) A closed loop through the Sycamore chip, simulating a 1D system. During each cycle, $\modG$ gates act first on the blue bonds, then on the black bonds. All other bonds remain idle during the dynamics.
\label{fig:snake}}
\end{figure}

NextGen programmable quantum simulators are designed with quantum computing applications as a major drive.
These applications happen to require many of the items of the above checklist.
The preparation of arbitrary computational-basis states and the capability for site-resolved read-out  are both key ingredients for quantum computing~\cite{DiVincenzoCriteria}, so it is fair to assume their availability on a NISQ device, up to small control and measurement errors.
Moreover, these devices are designed to implement quantum circuit elements that are typically one- and two-qubit gates, which in the quantum many-body language means on-site fields and nearest-neighbor interactions. While the selective realization (elimination) of short-ranged (long-ranged) couplings is an engineering challenge in all quantum computing platforms (currently addressed with varying degrees of accuracy in each one, and sure to see sustained effort in the future), it is fair to assume that on near-term digital quantum simulators crosstalk between distant qubits will be limited, and the dominant interactions will be between neighboring qubits.
Such finite-range interactions are much more suitable for MBL compared to the power-law decaying couplings native to many platforms~\cite{deroeckAvalanche}. 
Thus short-ranged interactions (requirement iii), site-resolved measurements (requirement vi) and tunable initial states (requirement v) are all at our disposal, within reasonable levels of approximation.
Moreover, as these devices enter the 50-to-200-qubit NISQ regime~\cite{Preskill2018} they can be safely regarded as legitimate quantum many-body systems (requirement i).

According to the checklist in Section~\ref{sec:ingredients}, The last two points to be addressed are (a) whether the coherence times are long enough, given the eponymous noise inherent to NISQ devices and (b) whether the devices can implement a kicked Ising drive similar to the one in Eq.~\eqref{eq:FloquetU}, with disorder in the Ising couplings, $J_{ij}$. 
While a universal fault-tolerant quantum computer can, of course, realize any drive with any set of couplings~\cite{Feynman1982, Aharonov1997}, present day NISQ devices may present obstructions due to their finite coherence time. Again, we are motivated by near-term applications that are immediately and \emph{naturally} realizable on these platforms (as opposed to universally and asymptotically). 
To address these points in a more specific way, we focus on Google's Sycamore processor for the remainder of this work. 
In Sec.~\ref{sec:model} we lay out the details of implementing the Floquet DTC as a quantum circuit with gates available on Sycamore, while in Sec.~\ref{sec:mbldtc} we map out the phase diagram of this circuit model and present several diagnostics of the MBL DTC phase. 
All of the analysis for now assumes an `ideal', i.e. decoherence-free realization; the analysis of noise which informs the coherence time is presented in Sec.~\ref{sec:noise}.

\subsection{Floquet DTC circuit on Sycamore\label{sec:model}}

We begin by noting that the Floquet unitary evolution operator for the canonical model of a DTC, Eq.~\eqref{eq:FloquetU}, can be naturally written as a sequence of gates when $H_\text{int}=0$, and when the $J_{ij}$ couplings are limited to nearest neighbors. 
We confine the dynamics to a one dimensional system, where the existence of MBL and thus of the DTC phase is on firmest ground~\cite{Imbrie2016, deroeckAvalanche}. In this case, one first acts with a layer of Ising gates $e^{-iJZZ}$ on the even bonds of the 1D subsystem, then a layer of Ising gates on the odd bonds, and then a layer of single-qubit $X$ rotations, $e^{-igX}$:
\begin{align}
    U_F &= e^{-i g \sum_i X_i} e^{-i \sum_i J_i Z_i Z_{i+1}} \nonumber\\
    & = \prod_i R^x_i(2g)  
    \prod_{i} e^{-i J_{2i-1} Z_{2i-1} Z_{2i}}
    \prod_{i} e^{-i J_{2i} Z_{2i} Z_{2i+1}}
\label{eq:FloqU_gates}
\end{align}
where $R^x_i(\alpha) = e^{-i\alpha X_i/2}$ is a single qubit $X$ rotation. This model has Ising symmetry and is exactly solvable, being mappable to free fermions. 
In this limit, the system is in the DTC phase (with period doubled dynamics and spontaneously broken Ising symmetry) as long as the average $J$ couplings obey~\cite{Khemani2016}
\begin{equation}
\left| \overline{J_i} - \frac{\pi}{4} \right| \leq  g-\frac{\pi}{4}
\label{eq:piSGphaseboundary}
\end{equation}
(one can take $g, \overline{J_i} \in [0,\pi/2]$ without loss of generality as the phase diagram repeats symmetrically outside this square). 
As mentioned earlier, the DTC phase persists for a finite region in parameter space surrounding $g=\frac{\pi}{2}$, even upon perturbing the drive in Eq.~\eqref{eq:FloqU_gates} with generic interactions to make the model non-integrable, as long as the disorder in $J_i$ is strong enough to stabilize MBL. 

On the Sycamore chip, a unitary evolution close to Eq.~\eqref{eq:FloqU_gates} can be straightforwardly implemented. Single qubit $X$ rotations $R^x_i$ are readily available~\cite{Google2019}. 
For the two-qubit interaction, the Sycamore device allows implementation of a continuously parameterized family of high fidelity gates of the form~\cite{DiabaticGates, gmonGates}
\begin{equation}
G_{1,2} = R_{1}^z(h_{a}) R_2^z(-h_{a}) 
\fsim_{1,2}(\theta,\phi) 
R_1^z(h_b) R_2^z(h_c) \;,
\label{eq:transmon_gate}
\end{equation}
where $R^z_i(\alpha) = e^{-i\alpha Z_i/2}$ is a single-qubit $Z$ rotation, the $h$ angles result from the frequency excursion of the single qubits during the interaction\footnote{There are only 3 independent angles because $Z_1+Z_2$ commutes with $\fsim$.}, and $\fsim$ is the `fermionic simulation' two-qubit gate~\cite{fSim2018},
\begin{equation}
\fsim_{1,2} (\theta,\phi) = e^{-i\frac{\theta}{2}(X_1X_2+Y_1Y_2) -i \phi \frac{Z_1-I}{2}\frac{Z_2-I}{2}} \;,
\label{eq:fsim}
\end{equation}
defined by an `iSWAP angle' $\theta$ and a `controlled-phase angle' $\phi$. 
The latter provides the crucial ingredient for the Floquet DTC unitary: the two-qubit Ising coupling $e^{-iJ ZZ}$, with the identification $J \equiv \phi/4$.

The remaining terms in Eq.~\eqref{eq:transmon_gate}, i.e. the iSWAP angle $\theta$ and the single-qubit $Z$ rotations (coming both from $\fsim$ and from the $h$ angles), represent deviations away from the solvable limit in Eq.~\eqref{eq:FloqU_gates}, but these deviations can be controlled and manipulated rather straightforwardly. 
Specifically, the angles $\theta_{ij}$, one for each coupler in the Sycamore chip, can be independently tuned to arbitrary values (including zero) within calibration accuracy. 
For the purpose of this paper we will sample each $\theta_{ij}$ out of a normal distribution with variable mean $\overline{\theta}$ and standard deviation $\Delta \theta = \pi/50$, representing gate calibration error of a few degrees ($\pi/50\ \text{rad} = 3.6^\circ$), a deliberately conservative upper bound. 
The `extra' single-qubit $Z$ rotations can also be tuned and cancelled ``by hand'' (within calibration accuracy) with active $Z$ rotations of appropriate angles on each qubit before and after each application of $G$, see Fig.~\ref{fig:snake}(a). 
The result is a modified gate 
\begin{align}
\modG_{i,j}
& = R_{i}^z(\delta h_a^{ij}) R_j^z(-\delta h_a^{ij}) e^{-\frac{i}{2} \theta_{ij} (X_iX_j + Y_iY_j)- \frac{i}{4} \phi_{ij} Z_i Z_j} \nonumber \\
& \qquad \times R_i^z(\delta h_b^{ij}) R_j^z(\delta h_c^{ij}) 
\label{eq:modified_G}
\end{align}
where the $\delta h$ are small residual rotation angles, taken to be normal random variables of standard deviation $\Delta h = \pi/50$. Note that the non-zero $\Delta h$, $\overline{\theta}$ and $ \Delta \theta$ make the model genuinely interacting and non-integrable; the $\Delta h$ terms also break the Ising symmetry. Both effects are necessary for a nontrivial demonstration of the stability of the phase. 
Thus, even as calibration errors continue to improve, these deviations can and should be deliberately included for a non-trivial demonstration of the phase.
We have explicitly verified by numerical diagonalization that $\Delta h = \Delta \theta = \pi/50$ is large enough to visibly break integrability even when $\overline{\theta}=0$. 

With the $\modG$ gate defined above, it is now straightforward to define our model Floquet circuit: 
\begin{equation}
U_F = \prod_i R^x_i(2g)  \prod_{ i} \modG_{2i-1,2i} \prod_{i} \modG_{2i,2i+1} \;,
\label{eq:FloquetCircuit}
\end{equation}
sketched in Fig.~\ref{fig:snake}(b). This represents a generically perturbed and non-integrable variant of the solvable model in Eq.~\eqref{eq:FloqU_gates}. 
Single-qubit rotations are widely and easily tunable on Sycamore, allowing for arbitrary values of the $\hat{x}$ rotation angle $2g$ (or equivalently the $\pi$ pulse imperfection $\epsilon = \pi-2g$). 
The two-qubit gates act, in turn, on the even and odd bonds along a one-dimensional path through Sycamore, such as the one sketched in Fig.~\ref{fig:snake}(c).
All the parameters specifying the individual $\modG_{ij}$ gates ($\phi_{ij}$, $\theta_{ij}$, $\delta h_{a,b,c}^{ij}$) are drawn randomly for each gate (one per spatial bond), but are \emph{time-independent}: all these choices are fixed once per realization, and then repeated in time so as to define an ideal time-periodic (Floquet) model\footnote{Any temporally random fluctuations and/or additional decoherence due to the execution of the active $Z$ rotations can be accounted for by increasing an effective `Pauli error rate'; we will return to them when we discuss the noise model in Section~\ref{sec:noise}.}. 
Again, we chose to use a one-dimensional path through Sycamore rather than the full 2D array of couplers in order to remain within the territory where MBL and the DTC phase are firmly established on theoretical grounds.
However we note the extreme flexibility of this platform in potentially choosing different geometries -- e.g. 1D paths of different lengths, with open or periodic boundary conditions, or 2D patches of various shapes -- all on the same chip, simply by selecting which couplers to activate and which to leave idle during the dynamics.

Having discussed the parameters $g, \theta_{ij}$, $\delta h_{a,b,c}^{ij}$ above, we now turn to the $\phi_{ij}$ angles, which set the strength of the $ZZ$ coupling and address the final requirement of Ising-even disorder. From an engineering perspective, two-qubit gates are generally more demanding than single qubit rotations:
each distinct gate acting on a given bond $\langle i,j\rangle$ must be calibrated individually~\cite{DiabaticGates}.
The phases $\phi_{ij}$ are thus drawn randomly from a \emph{discrete} set of $M$ values ($M\lesssim 10$ appears realistic in the near term), rather than a continuous distribution as is usually assumed in studies of MBL. This is because choosing gate parameters from a continuum would require one to calibrate each gate in the circuit for each distinct disorder realization, making the process highly impractical. 
In contrast, it is vastly easier to calibrate $M$ gates per bond at the beginning (so $\sim LM$ distinct gates in total), and then use these to generate a virtually infinite ($\sim M^L$) number of disorder realizations.

In this work we choose the discrete set of disordered couplings to be
\begin{equation}
\{\overline{\phi} + W\cos(\pi m/(M-1)):\ m = 0,\dots M-1\}\;,
\label{eq:discrete_disorder}
\end{equation}
where $\overline{\phi}$ sets the average coupling and $W$ the disorder strength.
The use of a nonlinear function ensures that there are incommensurate spacings between the different phases $\phi_{ij}$, thus limiting the effect of accidental resonances\footnote{Localization is expected to be stable even with discrete (rather than continuous) disorder, provided the number of values $M$ in the discrete set is large enough ($M=2$ is pathological)~\cite{Janarek2018}. Nonetheless, we note that discrete disorder falls outside the set of conditions required for a rigorous non-perturbative proof of MBL~\cite{Imbrie2016}, and may thus generate resonances that eventually destabilize localization. However, any such effects would appear on a parametrically long timescale, akin to concerns regarding the stability of MBL in higher dimensions or with power-law interactions of any power~\cite{deroeckAvalanche}. These open issues are beyond the purview of this work, and will be invisible at the system sizes and times accessible to near-term devices.}; the choice of $\cos(x)$ is otherwise arbitrary and is expected to yield generic results.
For specificity, in the following we fix the average controlled-phase angle to $\overline{\phi} = \pi$ corresponding to $\overline{J}=\pi/4$. This choice is at the center of the DTC phase in the non-interacting model, and allows for the widest range of rotation angles $g$ (cf. Eq.~\eqref{eq:piSGphaseboundary}). 
The disorder strength is set to $W = \pi/2$; this is fairly strong while also ensuring that all the $\phi$ angles are far from 0 (where the experimental implementation could be problematic in some cases~\cite{gmonGates}).  
Finally, we set $M=8$ based on numerical results obtained via full diagonalization of the Floquet unitary $U_F$ which indicate that that $M=8$ disorder values are sufficient to qualitatively replicate the continuous disorder ($M\to\infty$) case.

The quantum circuit so defined captures all the crucial aspects of the canonical Floquet DTC, Eq.~\eqref{eq:FloquetU}, in a ``Trotterized" form. It differs from the solvable limit, Eq.~\eqref{eq:FloqU_gates}, in specific ways:
the nonzero iSWAP angles $\theta_{ij}$ introduce interactions and make the model non-integrable;
the nonzero longitudinal fields, $\Delta h$, also add interactions and weakly break the Ising symmetry; 
and finally the disorder in the $\phi_{ij}$ couplings is discrete rather than continuous. 

In the following we confirm that these do \emph{not} destroy the DTC phase, as expected from its absolutely stable nature~\cite{vonKeyserlingk2016}. 
By varying $g$ and $\overline{\theta}$, with all other parameters fixed as described above, we obtain a phase diagram for the model circuit, shown in Fig.~\ref{fig:pd}. This was obtained by combining various phase diagnostics, discussed in the next section. 
It includes two MBL phases for sufficiently weak $\overline{\theta}$: a DTC phase near $g = \pi/2$ (corresponding to an imperfect $\pi$-flip), and a paramagnetic phase near $g = 0$. 
These are separated by a large thermal region, which expands as the interaction strength $\overline{\theta}$ is increased, eventually destroying both MBL phases for $\overline{\theta}\gtrsim \pi/8$.  The next section presents a detailed discussion of the diagnostics used to obtain this phase diagram and to detect the different phases in an experimental setting.


\subsection{Diagnostics of the MBL DTC phase}
\label{sec:mbldtc}

Nonequilibrium phases and phase transitions are understood as \emph{eigenstate phases}~\cite{lpqo, PekkerHilbertGlass, Bauer13, Parameswaran2017};
their theoretically sharpest diagnostics involve properties of the many-body eigenspectrum and of individual many-body eigenstates of the Floquet unitary $U_F$, which change in a singular manner across phase boundaries. 
While theoretically useful, these eigensystem diagnostics are not directly accessible to experiment, and their numerical exploration is limited to the small sizes amenable to exact diagonalization of $U_F$. Fortunately, these diagnostics translate to distinctive measurable signatures in dynamics from generic computational-basis initial states, that are both observable in experiment and accessible to numerics for much larger sizes.

We now present various eigenspectrum and dynamical diagnostics for identifying both MBL and the DTC order, which were used to derive a phase diagram for the model presented in the previous section.  
\vspace{6pt}

\begin{figure}
\centering
\includegraphics[width=\columnwidth]{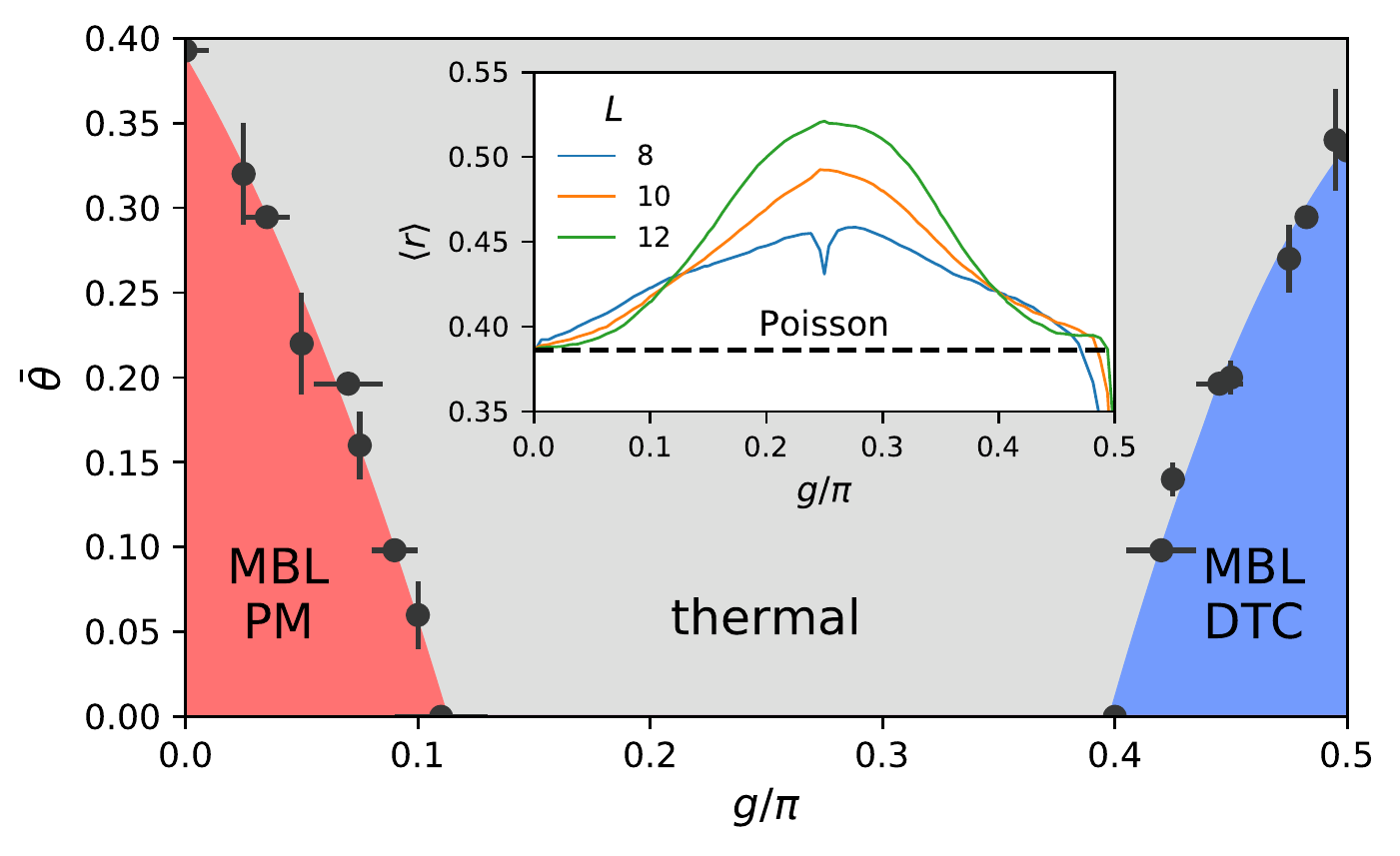}
\caption{Phase diagram of the circuit Eq.~\eqref{eq:FloquetCircuit} as a function of the pulse parameter $g$ and the average iSWAP angle $\overline{\theta}$. The phase boundary is based on finite-size crossing points (black dots) of the level spacing ratio, computed numerically for systems of $8\leq L \leq 12$ qubits.
Inset: level spacing ratio $\langle r \rangle$, Eq.~\eqref{eq:level_spacing_ratio}, vs $g$ on the $\overline{\theta}=0$ cut. $\langle r \rangle$ is averaged over eigenstates and over between 400 and 4000 realizations of disorder (depending on $L$).
\label{fig:pd}}
\end{figure}

\noindent{\bf Level Repulsion}
Many-body localization, aside from its dynamical signatures in the form of a persistent memory of initial conditions, is characterized by the absence of repulsion between quasienergy levels in the spectrum of $U_F$. 
The eigenvalues of $U_F$ are phases $\{e^{-i E_n}\}$; these can be used to obtain the quasienergies $\{E_n\}$, defined modulo $2\pi$. The statistics of quasienergy levels has been a powerful tool in the numerical study of MBL on finite systems, in particular the level-spacing ratio~\cite{Oganesyan2007}:
\begin{equation}
r = \frac{\min(\delta_n,\delta_{n+1})}{\max(\delta_n,\delta_{n+1})}
\label{eq:level_spacing_ratio}
\end{equation}
with $\delta_n = E_{n+1}-E_n$, the $n^\text{th}$ spacing between the quasienergies of $U_F$.
In an MBL phase, the value of $\langle r \rangle$ averaged over eigenstates and disorder realizations approaches the Poisson value 
$\langle r \rangle_\text{Poisson} \simeq 0.39$ with increasing system size, reflecting the lack of level repulsion that arises from localization.
In an ergodic phase it should instead approach the Gaussian unitary ensemble (GUE) value
$\langle r \rangle_\text{GUE} \simeq 0.60$, characteristic of random-matrix behavior~\cite{Atas2013}. 
Finite size scaling of this quantity across different cuts in parameter space was used to map out the phase diagram in Fig.~\ref{fig:pd}.
The inset displays one such cut, at $\overline{\theta}=0$, with two crossings separating the thermal phase ($\langle r \rangle$ increasing with $L$) from the two MBL phases ($\langle r \rangle$ decreasing with $L$).
Notice the dip below the Poisson value near $g=\pi/2$ is a finite-size effect due to the restoration of the Ising symmetry at $g = \pi/2$, where the $h$ fields are exactly `echoed out' over two periods.
\vspace{6pt}

\noindent{\bf Real-time oscillations}
The level spacing ratio distinguishes between MBL and thermal phases, but not between  different MBL phases. To do this, we need to consider specific features of the quantum order inherent in an MBL DTC. The hallmark of a DTC is spatiotemporal order: infinitely long-lived period-doubled oscillation of spins, in conjunction with long-range glassy order in space.
This is encoded in the behavior of a two-point correlation function~\cite{Khemani2016, vonKeyserlingk2016}
\begin{equation}
C_{ij}(n) = \langle Z_i(0) Z_j(n)\rangle \propto (-1)^n s_{ij}
\label{eq:spatiotemporal}
\end{equation}
at late times, where $n$ counts Floquet cycles and $s_{ij}$ encodes the ``glassy'' spatial order (i.e.\ is non-zero, but may have random sign as a function of $i$ and $j$).
This means memory of an initial glassy configuration is preserved forever, with the configuration itself flipped at every cycle.
Starting from a computational basis state $\ket{\psi(0)} = \ket{\boldsymbol \sigma}$ ($\boldsymbol\sigma\in\{0,1\}^L$), the statement in Eq.~\eqref{eq:spatiotemporal} simplifies to $\langle Z_j(n)\rangle \propto (-1)^{n} \langle Z_j(0)\rangle $: each spin gets flipped at every cycle, while maintaining a finite fraction of its initial (maximal) polarization. In contrast, an MBL paramagnet will retain memory of the initial configuration, but the spins do not get flipped.

\begin{figure*}
\centering
\includegraphics[width=\textwidth]{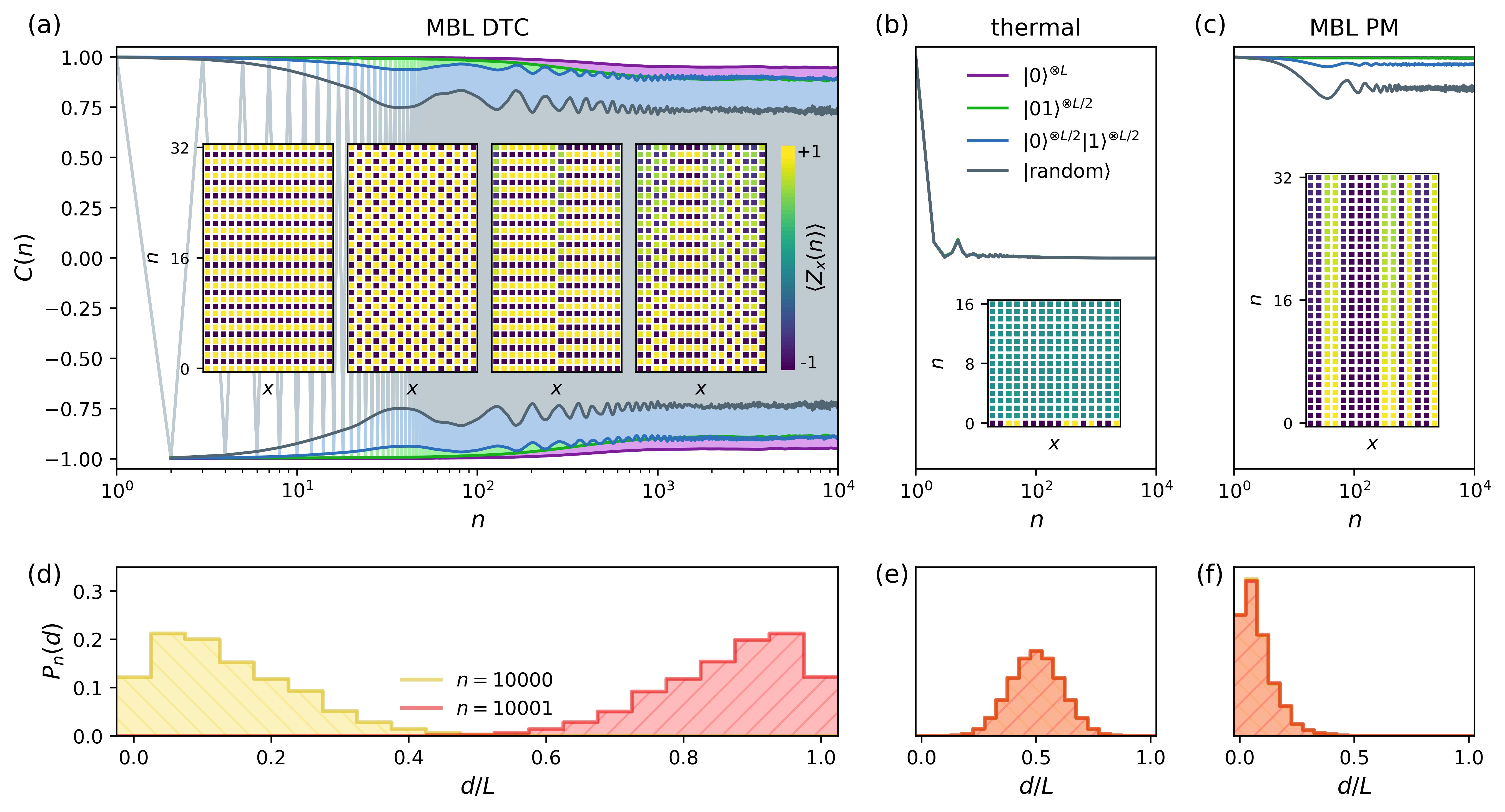}
\caption{\label{fig:unitary_dynamics} Dynamics of the ideal (noise-free) circuit in the MBL DTC ($g/\pi = 39/80$), thermal ($g/\pi = 19/80$), and MBL paramagnetic ($g/\pi = 1/80$) phases ($\overline{\theta}=0$).
(a-c) Position- and disorder-averaged temporal autocorrelator $C(n)$ starting from various initial bitstring states for $L=20$ qubits. In the DTC phase the envelopes at even and odd times are highlighted; the signal oscillates stroboscopically between these two envelopes (lighter curves).
Insets: space-time color plots of expectation values $\langle Z_x(n) \rangle$ for $L=16$ qubits. 
(d-f) Disorder-averaged probability distribution of the Hamming distance $d$ from the initial bitstring in two consecutive Floquet cycles at late time, $n = 10^4$, for $L=20$ qubits.
}
\end{figure*}

We perform exact numerical simulations of time-evolution (via sparse matrix-vector multiplication) under the circuit Eq.~\eqref{eq:FloquetCircuit} on systems of up to $L=22$ qubits starting from various computational basis states (ranging from polarized states to pseudorandom bitstrings). Representative plots for all three phases are shown in Fig.~\ref{fig:unitary_dynamics}(a-c) for $\overline{\theta}=0$ and one value of $g$ in each phase. 
We compute and plot $C(n) = \frac{1}{L} \sum_i \overline{C_{ii}(n)}$ which is the spatially resolved autocorrelator, Eq.~\eqref{eq:spatiotemporal}, averaged over all sites $i$ and over at least $10^3$ disorder realizations. 
In the DTC phase, all initial states show a persistent period doubled DTC signal $C(n)\propto (-1)^n$ up to at least $n_\text{max} = 10^4$ Floquet cycles (Fig.~\ref{fig:unitary_dynamics}(a)). 
In contrast, the MBL paramagnetic phase near $g=0$ shows a persistent signal $C(n)$, but at frequency $\omega=0$ rather than $\omega = \pi$ (Fig.~\ref{fig:unitary_dynamics}(c)). The large steady signal for a wide range of choices in initial states is a signature of MBL DTCs, which distinguishes them from prethermal DTCs. For example, a similar numerical simulation of autocorrelators in the trapped ion experiment sees strong state-to-state dependence, with $C(n)$ quickly decaying for most initial states~\cite{Khemani2019}. 
Finally, the behavior of both MBL phases should be contrasted with that of the thermal phase (Fig.~\ref{fig:unitary_dynamics}(b)) where the autocorrelator $C(n)$ quickly decays to zero for all initial states. 

The insets for panels (a-c) in Fig.~\ref{fig:unitary_dynamics} show space-time color plots of $\langle Z_i(t) \rangle$, visually depicting the oscillating glassy order in the MBL DTC, frozen memory in the MBL paramagnet, and rapid thermalization in the thermal phase. 
Importantly, measuring such \emph{site-resolved} space-time correlators for a \emph{wide range of initial states} is well within the existing capabilities of the Sycamore device. As discussed in Section~\ref{sec:ingredients}, such measurements are essential for a detection of the spatiotemporal order that defines the MBL DTC, and for distinguishing between MBL DTCs and prethermal variants. 
\vspace{6pt}

\begin{figure}
    \centering
    \includegraphics[width=\columnwidth]{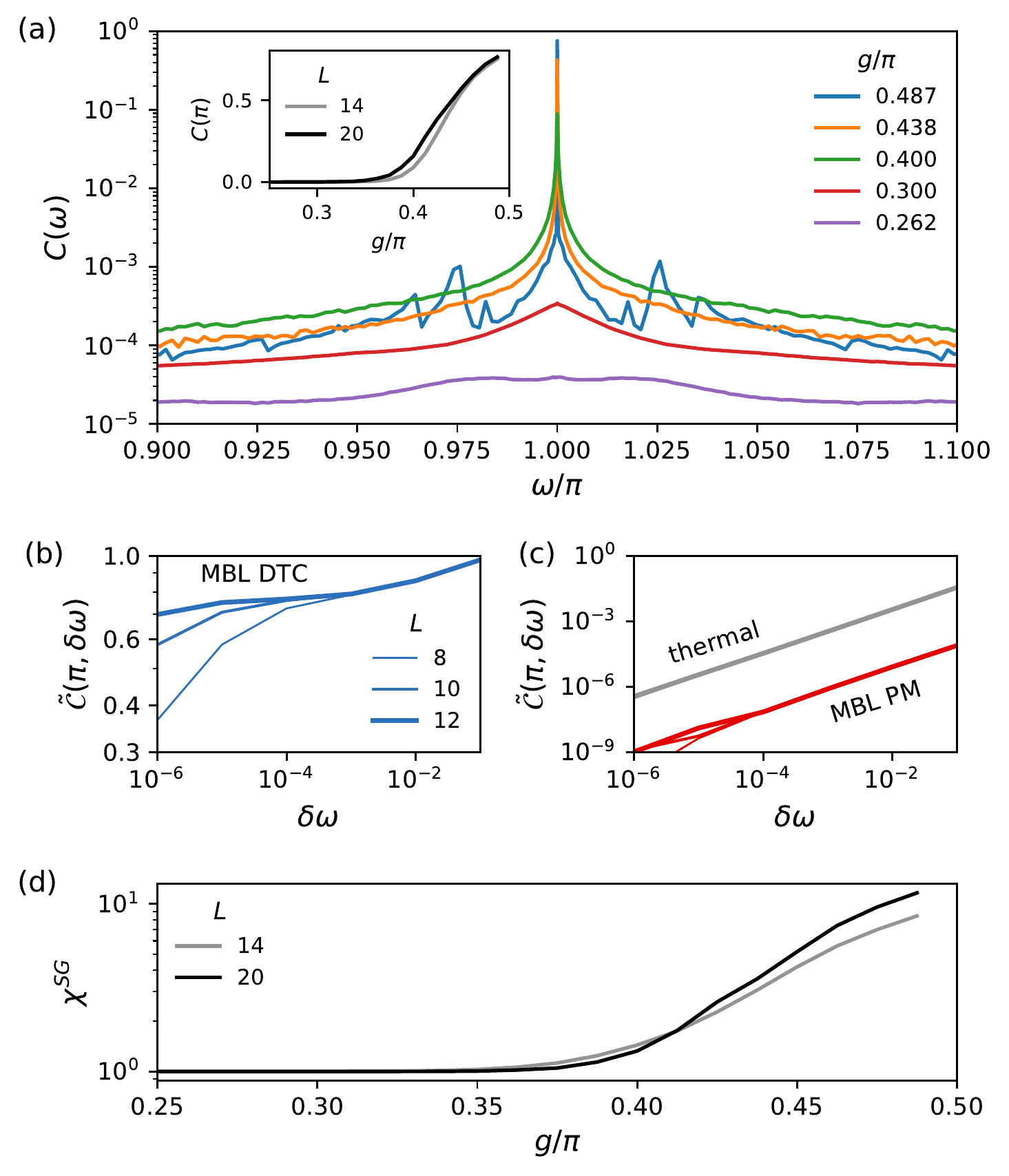}
    \caption{ \label{fig:chisg}
    Other diagnostics of DTC order.
    (a) Fourier transform of the temporal autocorrelator, $C(\omega)$, averaged over position and disorder, for several values of $g$ spanning the DTC and thermal phases. Data from dynamics simulations of $L=18$ qubits starting from a fixed bitstring state and evolving to $n_\text{max} = 10^4$ Floquet cycles.
    Inset: height of the $\omega = \pi$ peak as a function of $g$.
    (b,c) Spectral function $\mathscr{C}(\pi,\delta\omega)$ (see Eq.~\eqref{eq:integrated_spectral}) from exact diagonalization of $U_F$ on small sizes, averaged over disorder. The DTC phase develops a plateau for $\delta\omega\to 0$ corresponding to a delta-function $\pi$ peak in the Fourier response, while in the thermal and MBL PM phases we find $\mathscr{C}(\pi,\delta\omega)\sim\delta\omega$.
    (d) Spin glass order parameter $\chi^{SG}$ evaluated at late times, $n_\text{max}/2 \leq n \leq n_\text{max}$, from dynamics simulations as in (a). A crossing for increasing system size indicates a transition consistent with the phase boundary in Fig.~\ref{fig:pd} at $\overline{\theta}=0$.}
\end{figure}

\noindent{\bf Frequency-space peaks}
The real-time dynamics can also usefully be examined in frequency space, and used to probe how the DTC order melts and gives way to a thermal phase as the $\pi$ pulse imperfection $\epsilon = \pi-2g$ is increased~\cite{Khemani2016}.
Fig.~\ref{fig:chisg}(a) shows data obtained from dynamics simulations of $L = 14$ to 20 qubits at several values of the pulse parameter $g$ between $g = \pi/2$ (perfect $180^\circ$ pulse, center of the DTC phase) and $g = \pi/4$ (center of the thermal phase). The position- and disorder-averaged autocorrelator $C(\omega)$ (obtained from Fourier-transforming the real-time signal $C(n)$ collected out to $n_\text{max} = 10^4$) shows a peak at $\omega = \pi$ in the DTC phase, as expected; its height drops smoothly as one exits the phase (Fig.~\ref{fig:chisg}(a)).
While this is expected to sharpen with increasing system size, the finite-time limitation turns this into a smooth crossover (Fig.~\ref{fig:chisg}(a) inset). Such an analysis can, of course, also be done with experimentally measured dynamical signals.

Given that real-time dynamics simulations are inevitably limited to finite time $n$, a useful complementary perspective is achieved by examining spectral functions, where -- at the expense of more severe finite-size limitations -- we can effectively probe infinitely long times by a full diagonalization of the Floquet unitary $U_F$.
The period doubled behavior in Eq.~\eqref{eq:spatiotemporal} corresponds to a sharp delta-function peak at frequency $\omega = \pi$ in the spectral function
\begin{align}
\mathscr{C}_{ij} (\omega) 
& = \frac{1}{2^L} \sum_{\mu,\nu} \bra{\mu} Z_i \ket{\nu} \bra{\nu} Z_j \ket{\mu} \delta(E_\mu-E_\nu-\omega)
\label{eq:spectral}
\end{align}
where $\mu,\nu$ label the eigenstates of the Floquet unitary $U_F$ and $E_\mu$ are its quasienergies, i.e.\ $U_F \ket{\mu} = e^{-iE_\mu} \ket{\mu}$. This function represents a Fourier transform of the autocorrelator, Eq.~\eqref{eq:spatiotemporal}, over infinite stroboscopic times and averaged over all initial states.
It was used in conjunction with the level statistics to map the phase diagram in Fig.~\ref{fig:pd}, as described below. 

In a finite-size system, the spectral function $\mathscr{C}_{ij}(\omega)$ must be regularized by integrating over a finite frequency window $\delta \omega$,
\begin{align}
\tilde{\mathscr{C}}_{ij} (\omega,\delta\omega)
& \equiv \int_{\omega-\delta\omega}^{\omega+\delta\omega}d\omega' \mathscr{C}_{ij} (\omega') \;.
\label{eq:integrated_spectral}
\end{align}
A delta-function peak $\mathscr{C}_{ij} (\omega) \sim \delta(\omega-\pi)$ in the infinite size limit translates to a finite limit
$$
\lim_{\delta\omega\to0} \tilde{\mathscr{C}}_{ij} (\pi,\delta\omega) = \text{const.} \neq 0\;,
$$
as opposed to the generic non-DTC behavior $\tilde{\mathscr{C}}_{ij} (\omega,\delta\omega)\sim \delta\omega^\gamma, \gamma >0 $ as $\delta\omega\to0$.
Fig.~\ref{fig:chisg}(b,c) show numerical results for $\tilde{\mathscr{C}}(\omega = \pi, \delta\omega)$ at representative points in the three phases.
The onset of a plateau is clearly visible for increasing system size in the DTC phase, indicating the formation of a delta-function peak in $\mathscr{C}(\omega)$ at $\omega=\pi$.
Both the thermal and MBL paramagnetic phases instead obey the scaling $\tilde{\mathscr{C}} (\omega,\delta\omega)\sim \delta\omega \to 0$ as $\delta\omega\to0$.
\vspace{6pt}

\noindent{\bf Glassy spatial order.}
As discussed already, a key feature of the DTC phase is long-range \emph{spatial} `spin-glass' order which stems from spontaneously breaking an (emergent) Ising symmetry~\cite{Khemani2016, vonKeyserlingk2016}. 
This can be detected from long-range spatial correlation functions measured in the many-body eigenstates of the Floquet unitary (or, equivalently, from non-zero mutual information between distant subregions of the eigenstates~\cite{Else2016}). 
It can also be detected in dynamics through autocorrelators of the form Eq.~\eqref{eq:spatiotemporal}. 

Here we will use a classic diagnostic of spin-glasses related to the Edwards-Anderson order parameter~\cite{Edwards1975}:
\begin{equation}
\chi^{SG} = \frac{1}{L} \sum_{i,j} \bra{\psi} Z_i Z_j \ket{\psi}^2 \;.
\label{eq:chi_sg}
\end{equation}
This quantity is extensive in a phase with glassy order (where all $L^2$ items in the sum are finite); otherwise it is of order 1 (with only the $i=j$ contributions being significant). 
It can be examined in the many-body eigenstates of a Hamiltonian or of $U_F$~\cite{Kjall14, Khemani2016}, and its finite-size scaling provides yet another mechanism to deduce the phase diagram in Fig.~\ref{fig:pd}. 

Importantly, in a platform such as Sycamore with full spatial resolution, this quantity can also be examined dynamically starting from varied initial states. 
In Fig.~\ref{fig:chisg}(d), $\chi^{SG}$ (averaged over late times and disorder realizations) is plotted as a function of $g$ for $\overline{\theta}=0$,
and clearly shows a crossing with increasing system size, at a value of $g$ consistent with the phase boundary in Fig.~\ref{fig:pd}. Note that the effective system size probed on Sycamore can be easily varied by choosing which couplers to activate {\it i.e.} considering `snakes' of various lengths (cf. Fig.~\ref{fig:snake}). This presents a unique opportunity for \emph{experimentally} conducting finite-size scaling studies of the novel phase transition between the MBL and thermal phases, whose nature remains an active area of theoretical investigation. 
\vspace{6pt}

\noindent{\bf Hamming distance.} 
Finally, we present a diagnostic of spatiotemporal order that, while quite unusual from the point of view of many-body physics, is tailor-made for devices like Sycamore. 
The `quantum supremacy' experiment~\cite{Google2019} started with an initial bit string, time-evolved it under a random circuit, and then probed the output state by sampling its probability distribution over all bit-strings. 
We present a diagnostic for the different phases in our model that is in that vein, by considering the probability distribution of \emph{Hamming distances} between the initial and time-evolved states. 

Unlike ergodic dynamics, which quickly turns an initial bitstring state into a random state spread out over the entire computational basis, MBL prevents an initial state from veering too far from its initial condition.
This fact can be quantified by the Hamming distance $d$~\cite{Hauke2015, Smith2016},
which counts the minimum number of bit flips necessary to turn a bitstring $\boldsymbol\sigma \in \{0,1\}^L$ into another, $\boldsymbol\sigma^\prime$: for example
$d=0$ ($L$) only for identical (flipped) bitstrings, while typically $d=L/2$ between two random bitstrings.
Given a computational basis state $\ket{\psi(0)} = \ket{\boldsymbol{\sigma}}$ and its time evolution after $n$ Floquet cycles, $\ket{\psi(n)}$, we can define the Hamming distance distribution
\begin{equation}
P_n(d) = \bra{\psi(n)} \Pi_{\boldsymbol \sigma} (d) \ket{\psi(n)} \;,
\label{eq:hamming}
\end{equation}
where $\Pi_{\boldsymbol\sigma}(d)$ is the projector on bitstrings $\boldsymbol\sigma^\prime$ that are a Hamming distance $d$ away from $\boldsymbol\sigma$.
We note that the average of the Hamming distance distribution, 
$\overline{d} = \sum_d P_n(d) d$,
is information that can also be extracted from local expectation values of $Z$, since
$2\overline{d} = L - \sum_i (-1)^{\sigma_i} \langle Z_i(t)\rangle$;
in particular for the polarized initial state this becomes a global observable, the total magnetization $\sum_i Z_i$.
However the full distribution $P(d)$ requires measuring the probabilities of entire bitstrings -- a natural task for a programmable quantum simulator such as Sycamore that may instead be impractical or impossible on other platforms where such detailed read-out is unavailable.
While measuring the average $\overline{d}$ is enough to discriminate between MBL and ergodic phases, and a detailed measurement of the entire distribution $P_n(d)$ (particularly of its tails) would require considerably more sampling, it is nonetheless useful to have this capability.
Even a coarse estimate of the distribution's width would be informative about the size of the subset of Hilbert space explored by the initial state during the dynamics, which in turn relates to the localization length (i.e. the spatial extent of the local integrals of motion).

Fig.~\ref{fig:unitary_dynamics}(d-f) show data for the Hamming distance distribution $P_n(d)$ (Eq.~\eqref{eq:hamming}) in consecutive Floquet cycles at late times, $n_1=10^4$ and $n_2 = n_1+1$, in the three phases.
In the DTC phase (Fig.~\ref{fig:unitary_dynamics}(d)), $P_n(d)$ remains peaked near $d=0$ (the initial bitstring) at even $n$ and, symmetrically, near $d=L$ (the globally flipped initial bitstring) at odd $n$.
On the contrary, in the MBL paramagnet (Fig.~\ref{fig:unitary_dynamics}(f)) $P_n(d)$ remains peaked near $d=0$ at all times.
The behavior of both MBL phases should be contrasted with that of the thermal phase (Fig.~\ref{fig:unitary_dynamics}(e)), where the Hamming distance distribution quickly becomes peaked at $d=L/2$.

\section{Effect of Noise}
\label{sec:noise}

The discussion in the previous Section shows that the Sycamore device has, in principle, all the ingredients necessary to stabilize and detect a DTC phase.
We now address the important question of the robustness of the implementation and diagnostics to errors (in the form of noisy gates, environmental decoherence, and spurious time-dependence of the circuit parameters). 
These give a signal that will be decaying in time, in practice. As discussed below, estimates of current noise thresholds predict that the distinctive temporal signatures of DTC order should still be visible for multiple hundreds of driving periods. 
We emphasize again that spatial randomness is an inherent part of the DTC Floquet circuit, so small calibration errors between target gates and actual circuit elements are not a problem, provided these are reliably repeatable in time to give a Floquet circuit. 

We model noise by considering a one- and two-qubit depolarizing error model~\cite{NielsenChuangBook}, acting on the system's density matrix $\rho$ as
\begin{align}
\Phi_i^{(1q)} (\rho) & = (1-p_1)\rho + \frac{p_1}{3} \sum_{\alpha\neq 0} \sigma_{\alpha,i} \rho \sigma_{\alpha,i} \nonumber \\
\Phi_{ij}^{(2q)}(\rho) & = (1-p_2) \rho + \frac{p_2}{15} {\sum_{\alpha,\beta}}^\prime \sigma_{\alpha,i} \sigma_{\beta,j} \rho \sigma_{\alpha,i}\sigma_{\beta,j}
\label{eq:channels}
\end{align}
(the primed sum denotes $(\alpha,\beta)\neq(0,0)$).
Each single-qubit gate acting on a qubit $i$ is followed by an application of the channel $\Phi^{(1q)}_i$; each two-qubit gate on bond $(i,j)$ is followed by $\Phi^{(2q)}_{ij}$.
Conservative order-of-magnitude estimates for the depolarizing error rates with current technology~\cite{Google2019, DiabaticGates} are $p_1 \approx 10^{-3}$ and $p_2 \approx 10^{-2}$.
The additional errors introduced by the active single-qubit rotations in the definition of $\modG$ (Eq.~\eqref{eq:modified_G}) can be taken into account approximately by enhancing the values of $p_1$, $p_2$.
In the following we set $p_2 = p$, $p_1 = p/10$, and refer to the single parameter $p$ as the `Pauli error rate' unless otherwise specified. 

The channels Eq.~\eqref{eq:channels} subsume the effect of fairly generic experimental errors, e.g. environmental decoherence, temporally random fluctuations of gate parameters, etc.
In reality the errors may be anisotropic, e.g. $Z$ Pauli errors (phase-flip) may be more or less frequent than $X$ (bit-flip) errors.
While this issue can be completely neglected in ergodic circuits~\cite{Google2019}, where each qubit's Bloch sphere is quickly scrambled and the error model is made effectively isotropic, in this MBL setting this need not be true. 
Indeed, in structured evolutions that explore their Hilbert space unevenly, the effect of errors depends on the details of the circuit. Nonetheless, in the absence of more detailed device-specific error modeling, the depolarizing model is a reasonable choice in that it involves all Pauli errors.
We have additionally verified that our conclusions do not change qualitatively under a non-Pauli error model (the single-qubit amplitude-damping channel~\cite{NielsenChuangBook}), see Appendix~\ref{app:models}.

Quantum channels such as Eq.~\eqref{eq:channels} can be ``unraveled'' into stochastic unitary evolutions~\cite{Gisin1984, Dalibard1992}. 
Let us focus on the one-qubit channel $\Phi^{(1q)}_i$ for simplicity.
Its effect can be thought of as follows: 
after acting with each single-qubit gate $R^x_i$ from Eq.~\eqref{eq:FloquetCircuit}, the experimentalists toss a biased coin; with probability $p_1$, they apply an additional gate (``error'') drawn at random from $\{X_i, Y_i, Z_i\}$; otherwise they apply $I$ (i.e.\ they do nothing).
After $n$ cycles they get a pure state $\ket{\psi_{\mathbf{r}}(n)}$, where the label $\mathbf{r}$ keeps track of the error record, i.e.\ which error gates were applied, where and when during the entire evolution.
Iterating this stochastic process gives an ensemble of pure-state unitary evolutions (``quantum trajectories''~\cite{Brun2000}) $\{ \ket{\psi_{\mathbf r}(n)} \}$ that can be used to recover the density matrix $\rho(t)$ resulting from the \emph{real} noisy evolution:
\begin{equation}
\rho(n) \simeq \frac{1}{N_{\mathbf r}} \sum_{\mathbf r} \ket{\psi_{\mathbf r}(n)} \bra{\psi_{\mathbf r}(n)}\;,
\end{equation}
where $N_{\mathbf r}$ is the number of sampled trajectories (this becomes exact in the limit $N_{\mathbf r}\to\infty$). 
Thus at the expense of simulating multiple trajectories, one can evolve pure states instead of density matrices, greatly reducing the amount of memory needed for the computation.

\begin{figure}
    \centering
    \includegraphics[width=\columnwidth]{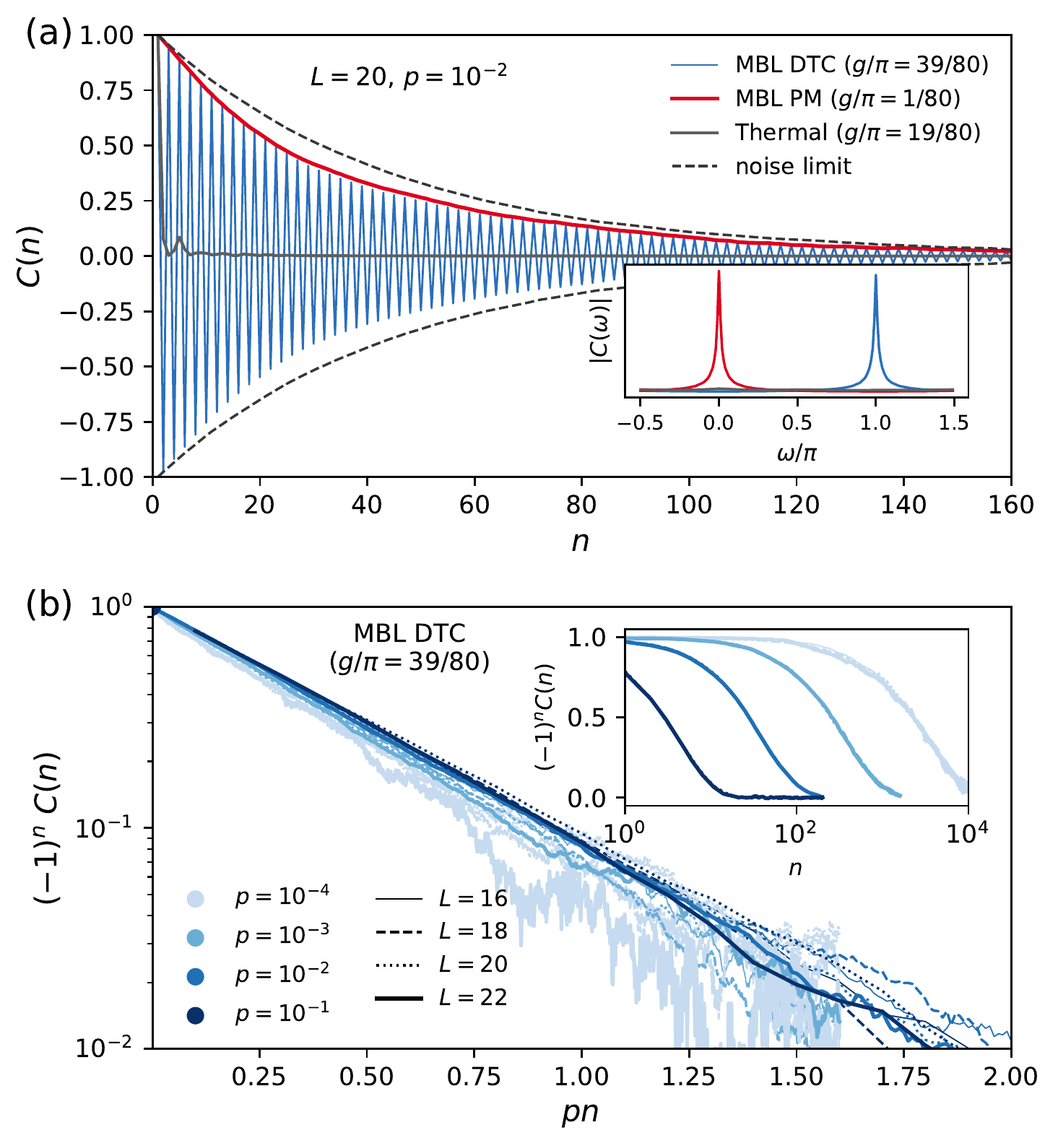}
    \caption{Noisy dynamics.
    (a) Time evolution of the spatially-averaged correlator $C(n)$ for a circuit with $L=20$ qubits in the presence of depolarizing noise ($p = 10^{-2}$) for the MBL DTC, MBL paramagnetic, and thermal phases, starting from a fixed bitstring state.
    The dashed line is the noise limit $e^{-\gamma t}$ (see main text).
    Inset: Fourier transforms of the signals show (broadened) peaks at $\omega = \pi$ for the MBL DTC and $\omega=0$ for the MBL PM.
    (b) DTC signal for different system sizes $L$ and error rates $p$. All curves in the main panel overlap within statistical error, showing that the signal does \emph{not} depend on $L$, and depends on $p$ only through the product $pn$, proportional to the number of accumulated errors \emph{per site}.
Inset: same data vs number of Floquet cycles $n$. Different system sizes are indistinguishable.
    }
    \label{fig:noise}
\end{figure}

Aside from their computational usefulness, quantum trajectories also offer a conceptually appealing view of the underlying error process. By unraveling a channel as outlined above, it is possible to think of the combined effect of all non-ideal processes taking place in the experiment as ``digital'', with discrete errors taking place at specific locations in spacetime during the circuit dynamics.
In the `quantum supremacy' experiment, Ref.~\onlinecite{Google2019}, it was argued that a single such digital error could completely randomize the output state: only ``error-free'' circuit realizations could contribute to the signal being measured in that work, hence its decay as $(1-p)^{Ln} \approx e^{-pLn}$ (for $p\ll 1$).
Therefore the signal's lifetime gets worse with increasing system size, $n^\star \sim 1/(pL)$.
This argument however need not hold for many-body localized (MBL) dynamics, where information propagates very slowly in space. 
It is plausible to expect in this case that a ``digital'' error at a given location will only affect observables in its vicinity, rather than completely randomize the output state.

This expectation is borne out by numerical simulations of quantum trajectories.
Given the depolarizing error model of Eq.~\eqref{eq:channels}, the autocorrelator $C_{ii}(n) = 
\langle Z_i (0) Z_i(n)\rangle$ inevitably decays in time.
Even under the ideal DTC circuit (with perfect $\pi$-pulse $\epsilon = 0$ and no $\theta_{ij}$ couplings) one can see that $Z$ operators decay exponentially:
$Z_i$ is invariant under the 2-qubit gates but decays under the subsequent error,  
$\Phi^{(2q)}_{ij}(Z_i) = (1-16p_2/15)Z_i$;
after two iterations of this (with its two neighbors), $Z_i$ picks up a minus sign under the $\pi$ pulse, followed by the decay under single-qubit noise $\Phi^{(1q)}_i (Z_i) = (1-4p_1/3)Z_i$.
Thus overall $Z_i \mapsto -e^{-\gamma} Z_i$ over one Floquet cycle, with 
\begin{equation}
\gamma = -\ln \left[ \left( 1-\frac{16}{15}p_2\right)^2 \left(1- \frac{4}{3}p_1\right)\right]
\label{eq:effective_noiserate}
\end{equation}
an effective decoherence rate.
Introducing non-ideal elements to the DTC drive ($\epsilon \neq 0$, $\theta_{ij}\neq 0$, etc) is not going to counter this decay; rather, it will generically include a (finite, transient) amount of `internal decoherence'.
The DTC signal is thus expected to be bounded by $\pm e^{-\gamma n}$.
The data in Fig.~\ref{fig:noise}(a) shows a DTC signal with amplitude close to the maximal level allowed by noise. 

Already with current hardware, this would yield a detectable DTC signal for hundreds of Floquet cycles.
Indeed, the measurement task consists of resolving the expectation $\langle Z_i(n) \rangle$ (which is small, $\sim e^{-\gamma n}$ at late times) of a binary variable with standard deviation $\sqrt{1-\langle Z_i(n)\rangle^2} \simeq 1$;
this requires repeating the same experiment $N_s\gg 1$ times, which is not a problem for the Sycamore device given its high speed of operation (for the `quantum supremacy' experiment~\cite{Google2019} $N_s = 10^6$ samples were obtained in a few minutes).
Equating the signal to the statistical noise floor then gives $e^{-\gamma n} \sim 1/\sqrt{N_s}$;
i.e. the signal can be resolved up to 
$n\leq n^\star \simeq \frac{1}{2\gamma}\ln(N_s)$.
Letting $p = 10^{-2}$ (a conservative estimate for the present technology) and $N_s = 10^6$ we obtain $n^\star = 303$ Floquet cycles.
We also note here that measurement is currently the lowest-fidelity process on Sycamore, with an average error rate of approximately $2.5\%$; however, this process happens only once per run, and thus its effect does not scale with the size or depth of the circuit\footnote{In practice, a rate of measurement error $p_\text{meas}$ increases the number of samples required, $N_s$, by a factor of $1/(1-p_\text{meas})^2$ -- i.e., by about $5\%$ for $p_\text{meas} = 2.5\%$.}.
The estimate above would improve logarithmically with the number of samples $N_s$, but most importantly it would improve linearly in the inverse Pauli error, which is set to see substantial improvements in the future.

Finally, some remarks about the statistics of sampling the DTC signal. 
First, measurement is destructive -- reading out the DTC signal at time $n$ arrests the evolution, which then needs to be started over for additional samples. Thus producing an experimental version of Fig.~\ref{fig:noise}(a) up to $n_\text{max}$ cycles would require a number of runs scaling with $n_\text{max}$. 
However, one could more economically extract robust evidence of spatiotemporal order from $O(1)$ time points, e.g. a snapshot at cycle $n$ (revealing spatial glassy order), one at $n+1$ (revealing the inversion of the glassy pattern), and one at $2n$ (revealing the stability of the original glassy pattern). 
Second, because we are ultimately interested in the quantity $\overline{\langle Z_i(n) \rangle}$, averaged not only over quantum measurements but also over independent disorder realizations of the circuit for each starting state, the $N_s = N_d N_q$ experimental runs will be divided in practice between $N_d$ disorder realizations and $N_q$ separate runs of each circuit for quantum averaging. Benchmarking and calibrating a given realization of $U_F$ is more experimentally demanding than multiple runs of the same circuit;  $N_q\sim O(10^4-10^6)$ and $N_d\sim O(10-100)$ seems feasible in the near-term, which should provide enough averaging to resolve the signal.

In contrast to the long-lived temporal signal in the DTC phase, the signal from a circuit in the ergodic phase decays within a few Floquet cycles. In the thermal phase, the signal lifetime is not limited by external noise but rather by \emph{internal} decoherence, i.e. by the system itself acting as a bath for the local observable~\cite{Nandkishore2015}, as shown in the error-free simulations of Fig.~\ref{fig:unitary_dynamics}(b).

Because the signal's lifetime in the DTC is only limited by external sources of error, future hardware improvements would directly translate to potentially much longer-lived realizations of the time crystal phase, as shown in Fig.~\ref{fig:noise}(b). 
Furthermore, the signal's decay rate does \emph{not} scale with size, suggesting that only errors in the vicinity of a given qubit cause damage to the local DTC signal.
To confirm this picture, we have also simulated the dynamics of a system where \emph{a single bond} $(i,i+1)$ is subject to decoherence, and we find that the local DTC signal $C_{jj}(n)$ decays as $e^{-n/\tau_j}$ with a time constant that diverges exponentially in the spatial distance from the faulty bond, $\tau_j \sim \exp {|j-(i+1/2)|/\xi}$, see Appendix~\ref{app:locality}. 
Thus when all bonds are noisy, by far the dominant source of decoherence for the signal at any site $j$ is the noise in its immediate vicinity, and the decay is to a very good approximation independent of $L$.

In sum: conservative estimates of noise levels suggest that Sycamore should already be able to observe a DTC signal for $\sim O(100)$ Floquet cycles, which is on par with what was observed in FirstGen experiments, but with significant improvements expected as the hardware continues to advance. Importantly, the signal decay time does not directly scale with system-size, so that the platform can be scaled up in size without a corresponding cost in experimental lifetime.


\section{Discussion and outlook}
\label{sec:discuss}

\subsection{Summary}

In this work we have considered the question: what does the dawning age of NISQ devices and programmable quantum simulators have in store for quantum many-body physics, focusing in particular on Google's Sycamore platform.
We have observed that, while these devices offer universal gate sets that can in principle simulate any quantum system, their limitation in coherence time practically favors certain simulation targets over others in the near term. Thus, when thinking of these devices as experimental platforms for many-body quantum mechanics, it is important to engage with their strengths and limitations, which are quite different from, and in some ways complementary to, those of the more traditional arenas for quantum many-body physics. 
This requires developing physical insight and intuition matching those needed in materials physics (regarding the choice of chemical compound, its synthesis, the selection and optimization of the experimental platform, and its theoretical modeling) or in cold-atomic systems (regarding the choice of atom or molecule, the cooling and loss suppression strategies, Hamiltonian engineering, and observable readout).

In the spirit of tailoring the application to what is most natural for the device in the near-term, we noted that unitary circuits implement various kinds of driven quantum evolutions more straightforwardly than they do time-independent Hamiltonians.
We have thus focused on out-of-equilibrium many-body phases in driven (Floquet) systems.
Specifically we have pointed to the Floquet discrete time crystal as a candidate well suited as a `physics forward' simulation task on Sycamore; this phase is simultaneously interesting as the first example of an intrinsically non-equilibrium many-body phase of matter, a good fit for Sycamore's capabilities, and not yet realized in any other experimental platform.
We have shown through detailed numerical simulations that the Floquet DTC can be stabilized on Sycamore over a range of realistic parameters, even under  conservative assumptions about gate calibration error, and that all facets of the DTC spatiotemporal order can be compellingly revealed using the device's extensive capability for initialization and site-resolved read-out.

We have also addressed the effects of noise and decoherence on detecting the DTC spatiotemporal order.
While all quantum simulators have to contend with the effects of environmental decoherence, the Sycamore platform has an edge insofar as the noise rates have been benchmarked with great care  (while a full characterization of the noise processes is an ongoing research effort~\cite{GoogleUpcoming}). 
This would make it easier, in practice, to disentangle the effects of `internal' and `external' decoherence upon observing a decaying signal in time. Further, the great control afforded by this platform could also permit the use of various `echo sequences' (such as one used in the NMR experiment, Ref.~\cite{BarrettExp, BarrettExp2}) to further separate the effects of internal and external decoherence.  
We note that the former is a matter of principle: if even in an ideal, noise-free model the signal is eventually destroyed by internal decoherence (i.e. quantum thermalization), then the system does not realize a DTC phase (this is true of all FirstGen DTC experiments).
On the other hand, if the signal's lifetime is 
limited by external decoherence (i.e. environmental noise and control errors), then this is an issue of engineering and, as such, will see sustained improvement with future hardware innovations.

Our proposal falls squarely in the latter category.
The signal lifetime, already in the hundreds of cycles with current technology, is predicted to steadily increase with hardware improvements.
The prospects for increasing the spatial size of the system are also promising. We have shown that the DTC order is sensitive to noise only locally, so that its lifetime is not negatively affected by increasing system size. The main constraint on the number of qubits thus becomes the geometry of the device.

\subsection{Directions for future research}

We conclude by mentioning interesting directions for future work along these lines.
A set of mild variations of the set-up proposed here can realize and probe a host of other interesting questions. 
Among these are prethermal time crystals~\cite{Else2017}, in particular in two dimensions. Experimental requirements are essentially identical to the ones we outlined, except of course for using all qubits and couplers on Sycamore's two-dimensional grid, rather than a one-dimensional subset of them.
Floquet symmetry-protected topological phases~\cite{Roy2017} are another natural target. These would require implementing a circuit that respects an Ising symmetry to a good approximation, and are thus a good target for future tests of high-precision many-body simulations.
Among two-dimensional nonequilibrium phases, the `anomalous' (or `chiral') Floquet insulator~\cite{Rudner2013, Po2016, Nathan2019} is another interesting target for simulation.
This phase, where an MBL bulk coexists with quantized, chiral information flow at the edge, would also be realizable as a quantum circuit within Sycamore's gate set. Specifically, its circuit implementation would consist of five steps: four of them are given by near-\textsf{SWAP} gates (i.e. angles $\phi \simeq \pi$, $\theta\simeq\pi/2$ in Sycamore's two-qubit gate set, with tolerance for sufficiently small imperfections), and the fifth is given by single-qubit disorder (e.g. $Z$ rotations by a site-dependent angle). The need for disorder only in single-qubit gates makes this particularly easy for Sycamore, as disorder realizations can be generated without additional calibration of two-qubit gates. Thus the gate set poses no problem. What may require further technological progress is size: the chiral Floquet insulator, and its signature quantized transport of quantum information at the edge, requires a clear demarcation between bulk and boundary, with states on distinct edges not interacting with one another. This may be out of reach with the current $\sim 6 \times 8$-sized device. A precise determination of requirements is a task for future research, as is the design of scalable protocols to measure the quantized flow of information within accessible coherence times.

Separately, quantum circuits are increasingly being studied as toy models for exploring a host of foundational questions in quantum statistical mechanics ranging from quantum chaos~\cite{Brown:2012aa, Hosur2016, Nahum2018, Curt2018, Gharibyan2018, Khemani:2018aa, ChanChalkerChaos, ProsenChaos} to the dynamics of quantum entanglement~\cite{Nahum2017, Curt2018, Rakovszky2018} to the emergence of hydrodynamics~\cite{kvh, Curt2018}.
Exploring some of these issues experimentally could have transformational impact on our understanding. 

Finally, a direction we leave for future study is that of estimating the classical computing resources needed to simulate the proposed circuits. Circuits implemented on a specific hardware platform in the presence of finite errors require careful estimates of classical computational resources. In general, however, we note there are no efficient classical algorithms for exploring the entire phase diagram in Fig.~\ref{fig:pd}.  
Indeed the nature of MBL-to-thermal phase transition is still a largely open question, in no small part because of severe finite-size effects plaguing numerical explorations~\cite{Pal2010, Luitz2015, CLO, KhemaniHuse2017, Vidmar2019, Abanin2019, Panda_2020, Sierant2020}. 
Experiments on analog quantum simulator platforms have already investigated many interesting features of the MBL phase ~\cite{Schreiber2015, KondovMBL, BlochMBL, BordiaSlowRelaxation, BlochSlowDynamics, GreinerMBLEntanglement, RispoliQuantumCritical, GoogleMBL}; the increased flexibility of digital platforms such as Sycamore may, in addition, enable experimental finite-size scaling studies of the MBL-to-thermal phase transition (cf. Fig.~\ref{fig:chisg}(d)), potentially reaching much larger sizes than existing numerical studies, which could lend important insights to some of these open questions. 

{\it Acknowledgments---} We thank Rahul Nandkishore and Siddharth Parameswaran for discussions. This work was supported with funding from the Defense Advanced Research Projects Agency (DARPA) via the DRINQS program. The views, opinions and/or findings expressed are those of the authors and should not be interpreted as representing the official views or policies of the Department of Defense or the U.S. Government.
MI was funded in part by the Gordon and Betty Moore Foundation’s EPiQS Initiative through Grant GBMF4302 and GBMF8686.
Computations were carried out on Stanford Research Computing Center's Sherlock cluster.
The work was in part supported by the Deutsche Forschungsgemeinschaft  through the cluster of excellence ct.qmat (EXC 2147, project-id 390858490).

\appendix

\section{Necessity of Ising-even disorder\label{app:echo_out}}

\begin{figure}
    \centering
    \includegraphics[width=\columnwidth]{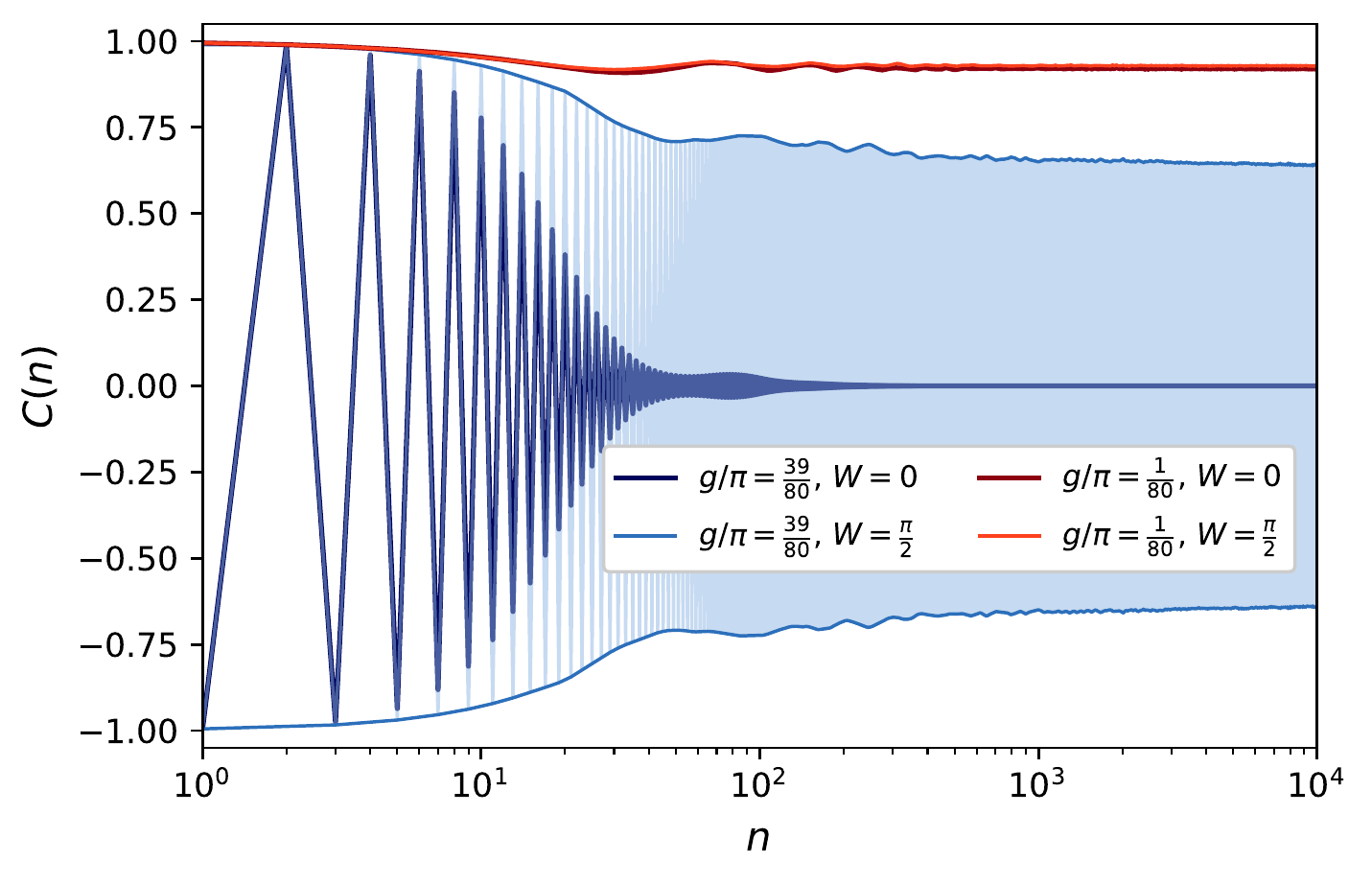}
    \caption{Temporal autocorrelator $C(n)$ for $L = 16$ qubits, averaged over position, 100 disorder realizations, and 100 initial states, with maximal disorder in the Ising-odd $h$ fields ($h \in [0,2\pi]$), and with disorder $W$ in the Ising-even $\phi$ angles.
    }
    \label{fig:echo_out}
\end{figure}

Here we explain why stabilizing an MBL DTC phase in the model Eq.~\eqref{eq:FloquetU} requires having disorder in the Ising-even couplings $J_{ij}Z_i Z_J$, whereas disorder in the longitudinal fields $h_iZ_i$ is insufficient. 
Considering the case $\theta_{ij}=0$ for simplicity (small non-zero values do not qualitatively change the argument), the time evolution over two consecutive periods is given by 
\begin{equation}
U_F^2 = P_{2g} e^{-iH_z[\mathbf{J},\mathbf{h}]} P_{2g} e^{-i H_z[\mathbf{J},\mathbf{h}]} \;,
\label{eq:twoperiods}
\end{equation}
where $H_z[\mathbf{J},\mathbf{h}] \equiv \sum_i J_i Z_iZ_{i+1} + h_i Z_i$ and $P_{2g} \equiv \prod_i R_i^x(2g) = \prod_i e^{-igX_i}$ is the imperfect $\pi$-flip, with $2g \equiv \pi-\epsilon$.
By using the fact that $Z_i$ anticommutes with the Ising symmetry $P_\pi = \prod_i X_i$, we rewrite Eq.~\eqref{eq:twoperiods} as
\begin{align*}
U_F^2 
& = P_{-\epsilon}\ e^{-i P_\pi H_z[\mathbf{J},\mathbf{h}] P_\pi} P_{-\epsilon}\ e^{-i H_z[\mathbf{J},\mathbf{h}]} \\
& = P_{-\epsilon}\ e^{-i H_z[\mathbf{J},-\mathbf{h}]} P_{-\epsilon}\ e^{-i H_z[\mathbf{J},\mathbf{h}]} \;.
\end{align*}
The crux of the argument is the fact that the fields $h_i$ have opposite signs in the two consecutive actions of $e^{-iH_z}$: to leading order in $\epsilon$, their effects cancel (``echo out''). 
To see this in more detail, we may write $U_F^2$ as
\begin{align*}
U_F^2 
& = P_{-2\epsilon}\ e^{-i P_\epsilon H_z[\mathbf{J},-\mathbf{h}] P_\epsilon^\dagger} e^{-i H_z[\mathbf{J},\mathbf{h}]} \\
& = P_{-2\epsilon}\ e^{-i P_\epsilon H_z[\mathbf{J},0] P_\epsilon^\dagger} \\
& \qquad \times e^{-i P_\epsilon H_z[0,-\mathbf{h}] P_\epsilon^\dagger}
e^{-i H_z[0,\mathbf{h}]}
e^{-i H_z[\mathbf{J},0]}\;;
\end{align*}
where we have decomposed $e^{-iH_z[\mathbf{J},\mathbf{h}]}$ into the (commuting) factors $e^{-i H_z[\mathbf{J},0]} e^{-i H_z[0, \mathbf{h}]}$.
Now if we take the $\mathbf{J}$ couplings to be clean, $J_i\equiv J$, the above expression can be rewritten by isolating the disordered part as
\begin{align*}
U_F^2 
& = U_\text{clean}^{(1)} \cdot
\prod_i e^{ih_i P_\epsilon Z_i P_\epsilon^\dagger} e^{-ih_i Z_i}  \cdot
U_\text{clean}^{(2)} \;.
\end{align*}
Straightforward algebra yields
\begin{align*}
e^{ih_i P_\epsilon Z_i P_\epsilon^\dagger} e^{-ih_i Z_i}
& = e^{-i\tilde{h}_i \hat{\mathbf{n}}_i \cdot {\boldsymbol\sigma}_i }
\end{align*}
where $\hat{\mathbf{n}}_i$ is a unit vector and $\tilde{h}_i$ obeys
\begin{equation}
\cos\tilde{h}_i = 1-\sin^2(h_i) \left(1-\cos\epsilon\right) \;,
\end{equation}
hence when $\epsilon \ll 1$ we have $\tilde{h}_i \approx \epsilon \sin h_i \ll 1$.
Thus the effective disorder strength in the fields $h_i$ is greatly reduced precisely in the regime where DTC order should be found (small $\epsilon$), posing a problem for the stabilization of the MBL DTC. Note that this is \emph{not} a problem at small $g$ ($\epsilon\approx \pi$), where disorder in the onsite fields does not get echoed out and can stabilize an MBL paramagnet.   
Numerical simulations of the model confirm this scenario, giving only an MBL paramagnetic phase (at sufficiently small $g$) and an ergodic phase in the rest of parameter space.

To illustrate this, we have performed dynamics simulations of the model realizable in Sycamore, Eq.~\ref{eq:FloquetCircuit}, 
with maximal disorder in the $h^{ij}_{a/b/c}$ angles (sampled uniformly from $[0,2\pi]$), both with and without disorder in the $\phi$ angles (again the identification between controlled-phase angles and Ising couplings is $\phi = 4J$).
We use the same discrete-disorder model as in the main text, with $M=8$ values, $\overline{\phi}=\pi$, and disorder strength $W$ set to either $W=\pi/2$ (as in the main text) or $W=0$. 
Finally we take $\overline{\theta} = 0$ and $\Delta\theta = \pi/50$.
The results are shown in Fig.~\ref{fig:echo_out}.
While the MBL PM phase ($g = \pi/80$) is fully stabilized by the $h$ fields, with negligible effect of $W$, the MBL DTC ($g = 39\pi/80$, i.e. $\epsilon = \pi/40$) requires $W \neq 0$.
In sum, in the absence of disorder in Ising-even interactions, disorder in the Ising-odd longitudinal fields $h_i$ is insufficient to stabilize the DTC phase. 

\begin{figure}
    \centering
    \includegraphics[width=\columnwidth]{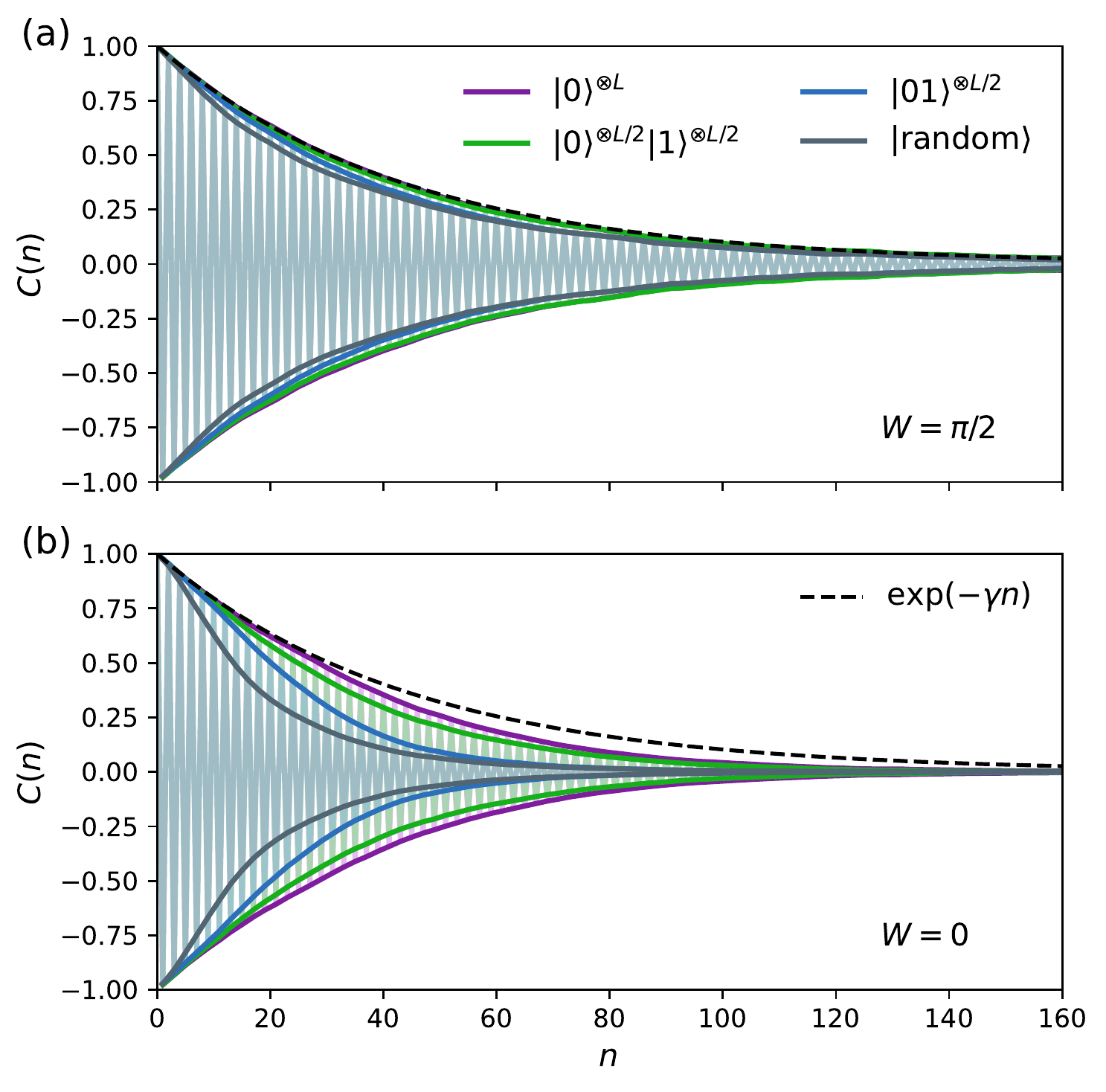}
    \caption{(a) Temporal autocorrelator $C(n)$ for $L = 20$ qubits in the presence of noise (rate $p = 10^{-2}$) in the MBL DTC phase ($g/\pi=39/80$, disorder in the controlled-phase angles $W = \pi/2$), for the four initial bitstring states indicated in the legend. 
    For each state, the data are averaged over position and at least $10^3$ combined realizations of disorder and noise (i.e. quantum trajectories). 
    The dashed line indicates the decoherence bound $e^{-\gamma n}$, also present in Fig.~\ref{fig:noise}(a), with $\gamma$ defined in Eq.~\eqref{eq:effective_noiserate}. The bitstring indicated by $\ket{{\sf random}}$ is $\ket{00101010100110111001}$.
    (b) Same simulation but without disorder in the controlled-phase angles, $W=0$ (with $\bar{\phi}=\pi$ as in (a)). 
    Other model parameters are $\Delta \theta = \bar{\theta} = \Delta h = \pi/50$.
    }
    \label{fig:noisypt}
\end{figure}

Finally, to show that this phenomenon would be visible even in the presence of external decoherence, we have repeated the analysis in the presence of an error rate $p = 10^{-2}$ (same as in Fig.~\ref{fig:noise}).
Results are displayed in Fig.~\ref{fig:noisypt}. 
For $W = \pi/2$ we have a genuine MBL DTC phase, showing very limited state-to-state variation and temporal autocorrelators $C(n)$ consistent with an $O(1)$ asymptotic value modulated by external decoherence, in line with the exponential envelope in Eq.~\eqref{eq:effective_noiserate}. 
On the contrary, in Fig.~\ref{fig:noisypt}(b) we turn off the disorder in the $\phi$ angles, setting $W=0$, and observe much stronger state-to-state fluctuations. Special ``low temperature'' states (such as the polarized one, or the one with only two domain walls) nearly saturate the decoherence envelope at early times; however they start to decay more quickly after a few tens of cycles -- at that point, one expects several bit flips have taken place due to the noise, and the states are progressively less ``special''. 
Typical (``high-temperature'') bitstrings, on the other hand, immediately decay faster than decoherence alone would dictate, indicating an intrinsic instability. 
This distinctive behavior is evident well within the coherence time of $\sim 100$ cycles.

\section{Effect of different noise models\label{app:models}}

\subsection{Control errors and decoherence}

In this work we have modeled the effects of noise and decoherence via a depolarizing channel. This is a justified assumption if the underlying dynamics is strongly scrambling, but not if it is highly structured, as in an MBL phase.
It is thus important to study the effects of different noise models.

To simplify the comparison of different models, we restrict to single-qubit decoherence channels, $\Phi^{(1q)}$. These are assumed to act after all gates, whether one- or two-qubit; so e.g. an application of $R^z_i$ is followed by the action of $\Phi^{(1q)}_i$, while an application of $G_{i,j}$ is followed by the action of $\Phi^{(1q)}_i \otimes \Phi^{(1q)}_j$.
We consider four families of quantum channels~\cite{NielsenChuangBook}, all parametrized by a rate $p\in[0,1]$ as follows:
\begin{enumerate}
    \item Depolarizing channel, $$ \Phi_i (\rho) = (1-p)\rho + \frac{p}{3} ( X_i \rho X_i + Y_i \rho Y_i + Z_i \rho Z_i) $$
    \item Bit-flip channel, $$ \Phi_i (\rho) = (1-p)\rho + p X_i \rho X_i $$
    \item Phase-flip channel, $$ \Phi_i (\rho) = (1-p)\rho + p Z_i  \rho Z_i$$
    \item Amplitude-damping channel, $$ \Phi_i (\rho) = A_i \rho A_i^\dagger + B_i \rho B_i^\dagger $$
    with
    $$
    A = \begin{pmatrix} 1 & 0 \\ 0 & \sqrt{1-p} \end{pmatrix},
    \quad
    B = \begin{pmatrix} 0 & \sqrt{p} \\ 0 & 0\end{pmatrix}\;.
    $$
\end{enumerate}
Unlike the depolarizing model, the other three act anisotropically on the Bloch sphere. 
Moreover, the amplitude-damping channel is not unital, i.e. does not preserve the maximally mixed state (it has $\ket{0} \bra{0}$ as its only fixed point).

\begin{figure}
    \centering
    \includegraphics[width=\columnwidth]{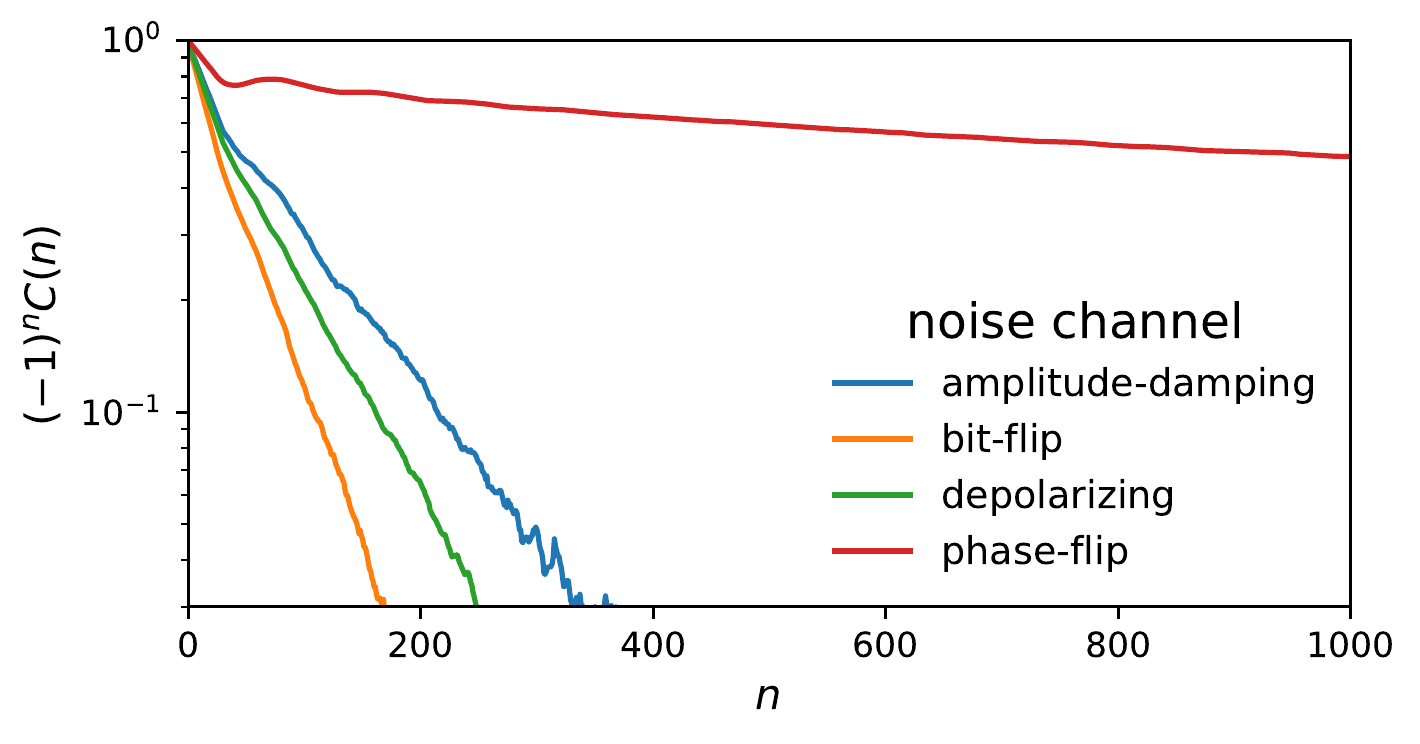}
    \caption{Effect of different noise models on the DTC signal (autocorrelator $C(n)$, averaged over position and $10^3$  disorder realizations) for $L=20$ qubits at $g/\pi = 39/80$. All noise channels are single-qubit and have rate $p=3\times 10^{-3}$ (see text).}
    \label{fig:noise_models}
\end{figure}

The results of simulations of the dynamics under these error models, using the same method described in the main text, are shown in Fig.~\ref{fig:noise_models}.
We find that the depolarizing, bit-flip and amplitude-damping channels have similar effects, causing an exponential decay with a rate close to $p$, within factors of order unity. 
On the contrary, the phase-flip error model causes a much slower decay (by over an order of magnitude in this case).

This behavior follows from the existence of local integrals of motion (lbits) in the MBL phase, which are nearly aligned with the $Z$ axis in these models. 
To the extent that the quantum jump operators $Z_i$ commute with the lbits, one can decouple the ideal dynamics (the Floquet unitary $U_F$) from the decoherence, 
$$
\rho\mapsto U_F^n \left[ \bigotimes_i \Phi_i^n(\rho) \right] (U_F^{\dagger})^n. 
$$
If the initial state is itself an eigenstate of the $Z_i$'s, it is immune to the phase-flip decoherence and one recovers the ideal dynamics.
Thus the effect of decoherence is suppressed by how closely aligned the lbits are with the $Z$ axis.
As the tilt is induced by the `transverse field' (pulse imperfection), one expects the rate of decoherence in this case to scale as $\sim |g|$ in the MBL PM phase and $\sim |g-\pi/2|$ in the MBL DTC. 
For the case of Fig.~\ref{fig:noise_models} this yields a lifetime enhancement of $\sim 40/\pi \simeq 13$, which is in line with the data.

Finally we note that the depolarizing noise is effectively a weighted average of bit-flip and phase-flip noise, and is thus intermediate between the two. 
The amplitude-damping noise behaves similarly as well.

\subsection{Measurement error}

Here we consider the effect of (possibly correlated) read-out errors on the various diagnostics studied in this paper.
We model the measurement error as a stochastic process where the outcome of a qubit state measurement (in the $Z$ computational basis) is randomly flipped, with probability $p_m$, away from its correct value. 
A realistic estimate for $p_m$ on Sycamore is $p_m\simeq 2.5\%$. 

We start by considering local observables, such as $C_{ii}(n) = \langle Z_i(n) Z_i(0)\rangle$. Assuming the initial state is a bit string $\ket{\psi} = \ket{\mathbf{s}}$ ($s_i\in\mathbb{Z}_2$) prepared perfectly, we have $C_{ii}(n) = s_i \langle Z_i(n)\rangle$; letting $A_\pm = \langle(1\pm Z_i(n))/2\rangle$ be the probability that qubit $i$ points up (down) at time $n$, the result of the noisy measurement process is $Z_i=+1$ with probability $A_+(1-p_m) + A_- p_m$, and $Z_i=-1$ with probability $A_+ p_m + A_- (1-p_m)$. 
In all, the estimate for $C_{ii}(n)$ with measurement error, $C_{ii}^\text{m.e.}(n)$, becomes
\begin{align}
    C_{ii}^\text{m.e.} (n) 
    & = A_+(1-2p_m) - A_-(1-2p_m) \nonumber \\
    & = (1-2p_m) C_{ii}(n) \;,
\end{align}
where we used that $C_{ii} (n) = A_+ - A_-$. 
This argument goes through for each site $i$, independent of any correlations in the measurement errors; hence averaging over position yields $C^\text{m.e.} = (1-2p_m) C(n)$, i.e. a damping by a time-independent overall prefactor. This lowers the signal's lifetime by a modest amount, but does not qualitatively change its behavior.

Correlations in read-out errors have an effect on quantities that specifically diagnose the spatial glassiness, such as the Edwards-Anderson (spin glass) order parameter, Eq.~\eqref{eq:chi_sg}.
Qubits on Sycamore are measured in groups of six via a frequency-multiplexing scheme~\cite{Google2019}, which could introduce correlations in the measurement errors. As we envision an effective one-dimensional system living on a path that zig-zags through Sycamore, such correlations in measurement errors will in general be non-local in the one-dimensional system. 
We consider an extreme scenario where the system's qubits are partitioned in groups $G_1, G_2, \dots $ and the measurement errors are \emph{perfectly correlated} within each group (and uncorrelated between groups): for each set $G_\alpha$, with probability $p_m$, \emph{all} qubits in $G_\alpha$ are measured incorrectly (i.e. flipped), otherwise they are all measured correctly. 
(One could study models with imperfect correlations in measurement errors and arrive at similar conclusions.)
The Edwards-Anderson order parameter is measured by first obtaining a quantum average of $s_{ij} \equiv \langle Z_i(n) Z_j(n)\rangle$, then computing $\chi^{SG} = \frac{1}{L} \sum_{i,j} s_{ij}^2$ and averaging the result over disorder realizations. 
We have that, for two qubits in the same group $G_\alpha$, $s_{ij}$ does not suffer any measurement error -- either both qubits flip, or neither does, leaving the product fixed. 
However for two qubits in distinct groups we have $s_{ij} \mapsto s_{ij} [(1-p_m)^2 + p_m^2 -2p_m(1-p_m)] = s_{ij} (1-2p_m)^2$.
Adding up all contributions, we find
\begin{align}
    \chi^{SG,\text{m.e.}} 
    & = \chi^{SG}_\text{diag} + (1-2p_m)^4 [\chi^{SG} - \chi^{SG}_\text{diag}]
    \label{eq:chi_sg_correlated}
\end{align}
where we introduced the ``diagonal'' sum
$$
\chi^{SG}_\text{diag} = \frac{1}{L} \sum_\alpha \sum_{i, j \in G_\alpha} s_{ij}^2 \;.
$$
The case of uncorrelated measurement errors is recovered by setting $G_\alpha \equiv \{i_\alpha\}$ (each qubit forms its own set), which gives $\chi_\text{diag}^{SG} = 1$ and thus 
\begin{align}
\chi^{SG, \text{m.e.}} 
& = 1 + (1-2p_m)^4 (\chi^{SG}-1) \;.
    \label{eq:chi_sg_uncorrelated}
\end{align}
Because all partial sums are positive, one has $\chi^{SG}_\text{diag} \geq 1$ in the presence of correlations.
It follows, by comparing Eq.~\eqref{eq:chi_sg_correlated} and \eqref{eq:chi_sg_uncorrelated}, that correlations in measurement errors in fact slightly \emph{enhance} the lifetime of $\chi^{SG}$ (compared to uncorrelated errors), and in any case do not qualitatively change its behavior.

\section{Locality of decoherence in MBL phases\label{app:locality}}

Here we discuss the effects of MBL on the propagation of decoherence in the system.
In a strongly ergodic system, an error anywhere in the system quickly randomizes the entire wavefunction (within a ballistic lightcone).
In an MBL system, on the other hand, an error at a given location has effects only within a \emph{logarithmic} ``lightcone''~\cite{Bardarson2012, Serbyn2013}, so for all practical purposes the effects of decoherence are local -- loss of coherence at a given site is dominantly the consequence of errors \emph{at that site} or in its immediate vicinity.

\begin{figure}
    \centering
    \includegraphics[width=\columnwidth]{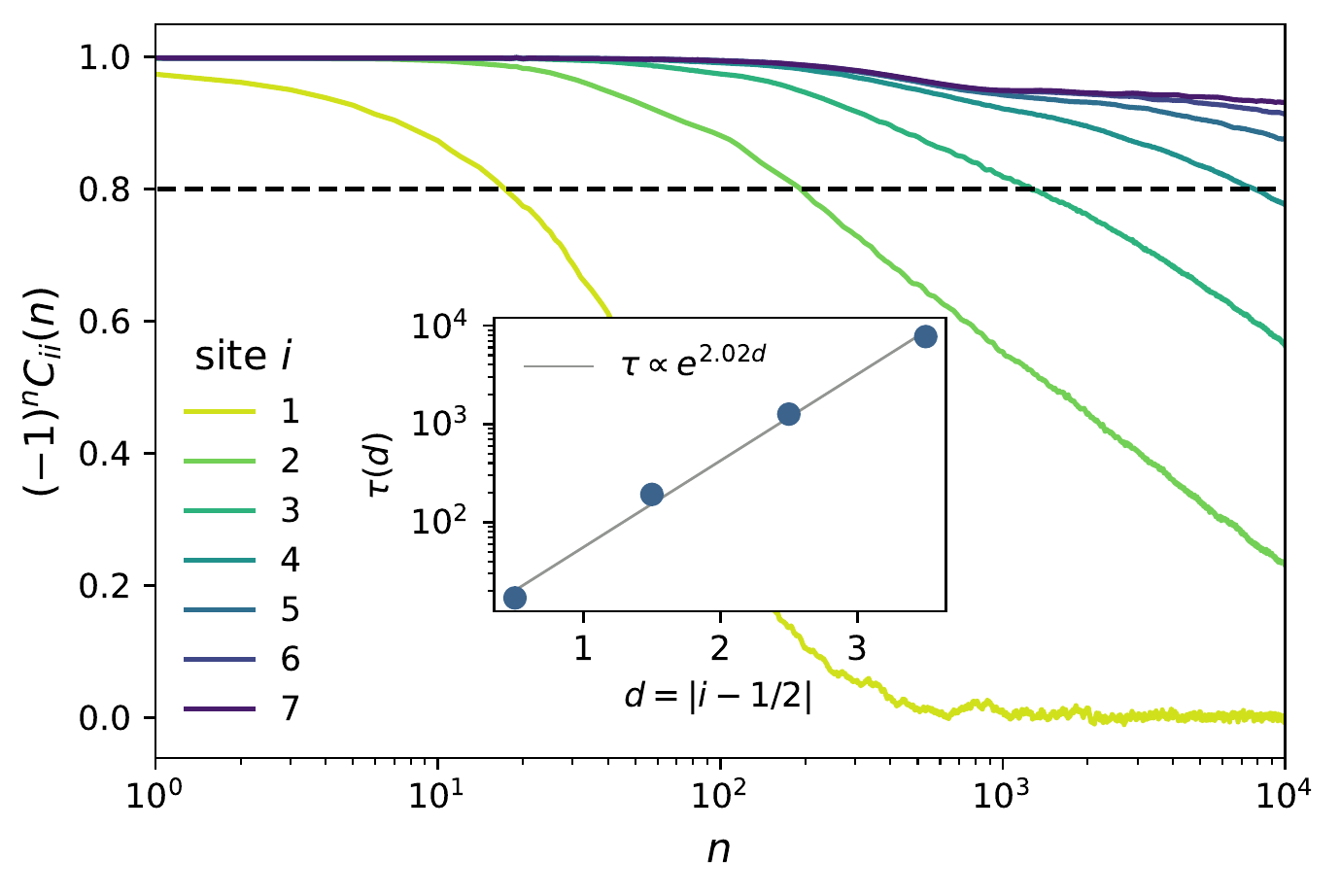}
    \caption{Numerical simulation of $L=16$ qubits in the DTC phase with depolarizing noise (as in Fig.~\ref{fig:noise}, with $p=0.01$) \emph{acting only on a single bond} (qubits $i=0,1$; boundary conditions are periodic), averaged over $10^4$ realizations of disorder. The dashed line represents an arbitrary threshold (0.8) used to extract a `decay time' $\tau$ for each site.
    Inset: DTC signal's decay time, $\tau(d)$, diverges exponentially in the distance $d$ from the noisy bond, consistent with the presence of exponentially localized integrals of motion in the MBL DTC phase.}
    \label{fig:noise_locality}
\end{figure}

We illustrate this point by simulating a qubit chain where the evolution is ideal and unitary everywhere, except for a single location in space.
For specificity we choose a bond (qubits $i=0, 1$); there, the same (one- and two- qubit) depolarizing noise model used in the main text acts at every time step. We then measure the DTC signal, i.e. the staggered autocorrelator $(-1)^n C_{ii}(n)$ at all qubits $i$.

The results are shown in Fig.~\ref{fig:noise_locality}.
We find that the decay time scale for qubit $i$ diverges as a function of its distance $d$ from the faulty bond as $\tau(d) \sim e^{d/\xi}$. 
This is consistent with the expectation for a system with exponentially localized lbits: each lbit is depolarized at a rate proportional to its overlap with the noisy sites, which in turn is set by the exponentially decaying envelope of the lbit, $\tau^z_i \sim \sum_j e^{-|i-j|/\xi} O_j$, where each $O_j$ is supported around site $j$.

As a consequence of this, loss of DTC signal at a given position is chiefly the result of errors at or near that position, even after a long time.
Thus the overall lifetime of the DTC signal is approximately independent of system size, as seen in Fig.~\ref{fig:noise}.

\bibliography{sycamoreDTC}

\end{document}